\documentclass[preprint,12pt, amsmath,amssymb,]{elsarticle}

\usepackage{amssymb}
\usepackage{lscape}
\usepackage{booktabs}
\usepackage[flushleft]{threeparttable}
\usepackage{amsmath}
\usepackage{xcolor}

\journal{Sensor and Actuators A: Physical}

\begin{document}

\begin{frontmatter}



\title{Hysteresis Compensation in Temperature Response of Fiber Bragg Grating Thermometers Using Dynamic Regression}


\author[inst1]{Zeeshan Ahmed}

\affiliation[inst1]{organization={National Institute of Standards and Technology, Physical Measurement Laboratory, Sensor Science Division},
            addressline={100 Bureau Drive}, 
            city={Gaithersburg},
            postcode={20899}, 
            state={MD},
            country={USA}}


\begin{abstract}
In recent years there has been considerable interest in using photonic thermometers such as Fiber Bragg grating (FBG) and silicon ring resonators as an alternative technology to resistance-based legacy thermometers. Although FBG thermometers have been commercially available for decades their metrological performance remains poorly understood, hindered in part by complex behavior at elevated temperatures. In this study we systematically examine the temporal evolution of the temperature response of 14 sensors that were repeatedly cycled between 233 K and 393 K. Data exploration and modelling indicate the need to account for serial-correlation in model selection. Utilizing the coupled-mode theory treatment of FBG to guide feature selection we evaluate various calibration models. Our results indicates that a dynamic regression model can effectively reduce measurement uncertainty due to hysteresis by up to $\approx  70 \%$ . 
\end{abstract}




\end{frontmatter}


\section{Introduction}
\label{sec:sample1}
Temperature measurements encompass almost every aspect of modern life ranging from advanced manufacturing to health screening constituting a multi-billion-dollar enterprise that is expected to continue growing as the use of temperature sensors proliferate \cite{Xu:14} (and refs within). Many thermometry techniques have been developed to meet the varied needs of the user community including resistance-based devices, such as thermistors and platinum resistance thermometers, as well as other sensing modalities including thermocouples, diodes and florescent probes. Standardized manufacturing of metal and semi-conductor based devices ensures an acceptable uncertainty (100 mK to few kelvin) over a given temperature range using nominal coefficients. Tighter uncertainty performance ($< 100$ mK) requires time consuming calibrations of each individual sensor \cite{Preston,MPK2019,Mangum_2002}. 
\par The size of the temperature sensor market is a powerful motivator for developing novel technologies targeted towards meeting present and future measurement needs. The existing metrology infrastructure and user expectations of minimum uncertainty metrics along with C-SWaP (cost, size, weight and power) performance requirements represent a formidable barrier to wide-spread adoption of any new temperature measurement technology \cite{AHMED_IMEKO}. As such, any emerging technology is expected to not only provide a novel utility but be backwards compatible with existing infrastructure. Photonic thermometers due to their small size, excellent thermal conductivity and compatibility with telecom infrastructure are expected to meet or exceed user demands\cite{AHMED_IMEKO}. In recent years the photonic thermometry community has largely focused on exploration of novel materials (e.g. silicon\cite{Xu:14,Krenek_ring_resonator,Sergey_ring,Yije_Fano}, silicon nitride\cite{Silicon_nitride}, diamond, etc\cite{Smith:21}), device configurations (Bragg waveguides\cite{WBG_optics_letter}, ring resonators\cite{Xu:14,Krenek_ring_resonator}, photonic crystal cavities\cite{PhCC} etc) and instrumentation to widen the application window of photonic thermometers beyond metrology labs\cite{Krenek_ring_resonator}. Until recently, systematic examination of temperature response of these devices including detailed characterization of measurement uncertainties has been lacking.  Several authors have examined the behavior of type-I and type-II fiber Bragg gratings (FBG)\footnote{grating types refer to photo-sensitivity mechanism used in writing of the grating. Type 1 rely on UV inscription in photosensitive fibers while Type II gratings are written using localized "damage" caused by two photon absorption} sensors at high temperatures and found that the sensors undergo significant hysteresis that is dependent upon both temperature and duration of excursion \cite{Ahmed_NCSLI,Sergey_FBG,PTB_FBG}. These results are broadly in agreement with earlier research on FBG fabrication processes that suggests the fabrication process creates shallow trap states in the bandgap that are "erased" at temperatures higher than 450 K\cite{Erdogan_decay,FBG_decay, FBGthesis,Munko_2016}. In addition, thermally driven ion migration between the fiber core and cladding, glass transition driven stress-strain changes in the fiber, crystallization of $\alpha$-quartz phase, grating erasure at elevated temperatures and mode mixing are suspected to contribute to measurement uncertainty\cite{FBG_decay,FBGthesis}. At temperatures below 450 K, mechanistic details of thermal hysteresis are unknown\cite{Erdogan_decay,Grattan}. Understanding the mechanism responsible for the hysteresis and quantifying its time-dependent impact on measurement uncertainties are the next steps in the development of FBG thermometers. In this study, we take a physics-informed approach to modeling hysteresis induced changes in the temperature response of FBG sensor. We rely on methods of machine learning and time series forecasting to develop a practical model that can be cost-effectively deployed in industrial setting. We note that elucidation of mechanistic details of hysteresis process i.e. specific changes to the chemical potential or bandgap of the sensor is beyond scope of this study.

\section{Experimental}
In this study we have utilized commercially available FBG acquired from five different vendors. One set of sensor were coated with a protective layer of polyimide while another set of sensors were coated with an acrylic layer. All other sensors were acquired without the polymeric coating. The fibers were stored in a humidity controlled environment (20$\%$ Relative Humidity) prior to use. Each fiber was cleaved such as to leave 2 mm of excess fiber on one side of the sensor, with the other side, 0.5 m long, terminated in a fiber optic coupler. Unless noted otherwise, the sensor was then guided through a T-coupler into a glass tube. The active sensing area of the sensor, at the bottom of the glass tube, was placed inside a through-hole opening (200 $\mu $m  dia.) of a small copper cylinder. The copper housing provides a strain-free mechanism for anchoring the loose fiber end whilst simultaneously providing a large thermal mass to ensure the the sensor remains in steady equilibrium. The glass tube was then continuously flushed with free-flowing Ar gas to prevent moisture condensation at temperatures below 283 K. \par The interrogation system has been described in detail elsewhere\cite{Ahmed_NCSLI}. Briefly, the assembled FBG thermometer was placed in a cylindrical Aluminum block (25 mm diameter, 170 mm length). The cylinder has two 150 mm long blind holes (2.5 mm and 6.5 mm diameter) for accommodating a calibrated thermister or platinum resistance thermometer and the assembled FBG sensor, respectively. The calibrated thermometer's uncertainties  over the temperature range of 233 K to 393 K are below 10 mK. The Aluminum block is placed inside the dry-well calibrator (Fluke 9170 ) whose temperature is controlled by software written in LabVIEW that cycles the temperature between 233 K to 393 K at preset intervals intervals (typically 5 K). Once the set temperature is achieved, the program allows for an equalisation period (20 mins unless noted otherwise). Following the equalization period the laser (New Focus TLB-6700\footnote{Any mention of commercial products is for informational purposes only; it does not imply recommendation or
endorsement by NIST.}) is scanned over a $\leq$ 2 nm window around the Bragg reflection peak. A small amount of laser power was immediately picked up from the laser output for wavelength monitoring (HighFinesse WS/7) while the rest, after passing through the photonic device via an optical circulator (ThorLabs CIR1550PM-FC), was detected by a large sensing-area power meter (Newport, model 1936-R). Consecutive scans were recorded at each temperature and each sensor was thermally cycled at least three times in each run unless noted otherwise. The recorded data was fitted using a cubic spline to extract peak center, peak height, and peak width as a function of temperature. The assembled dataset contains temperature response of 22 sensors including three quarter-phase gratings and two regenerated-FBG. The r-FBG's response is not included in the analysis presented below. Six other sensors were eliminated from final consideration due to insufficient number of thermal cycles (one or less cycles were successfully collected). Features recorded against temperature include date and time of the measurement, peak center, peak height, full width at half max,  area, kurtosis, sensor/grating type, coating, vendor (composite variable standing in for fabrication process variability), time, laser power and experiment type (experiments where number of consecutive scans is 100 or greater are referred to "annealing" as these experiments were designed to detect any slow relaxation process that might be occurring following temperature step). Data exploration was carried out using standard python\cite{python3} libraries (pandas\cite{pandas}, seaborn\cite{Seaborn}, matplotlib\cite{matplotlib}) while sklearn sci-kit\cite{scikit-learn} and statmodels\cite{statsmodels} libraries were used for data modeling. A brief discussion of the exploratory data analysis and methodology employed for data modeling is included in the supplemental.

\section{Results and Discussion}
\begin{figure}[htbp!]
\centering\includegraphics[width=12 cm]{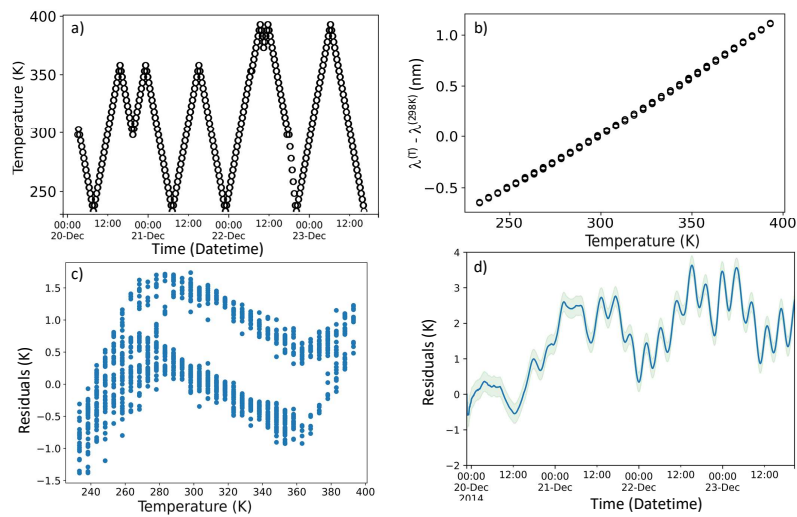}
\caption{a) Measured temperature cycling profile for sensor S3. b) Measured wavelength detunning vs temperature shows strong quasi-linear dependence that can be modeled as a quadratic function. c) residual of a quadratic fit shows significant departures from normal distribution due to hysteresis. d) time dependence of the residuals from the quadractic fit shown in solid line clearly exhibits a linear increasing trend. The shaded region marks the confidence interval for one-step prediction of AutoRegressive Integrated Moving Average (ARIMA) models trained on the calibration ramp (see discussion for details.}
\end{figure}
We explored the complete dataset for possible correlations between temperature and sensor features (see supplemental for details). Exploratory data analysis indicates that besides peak center, which is strongly correlated with temperature, a multitude of features show some degree of correlation with temperature and could be useful in constructing a temperature inference model. Using all of these features together, however would be imprudent. To whittle down the number of candidate features we note that the temperature response of the FBG (and any photonic thermometer in general) relies on the thermo-optic coefficient to transduce temperature changes into the frequency changes\cite{AHMED_IMEKO} (and references within). We therefore, use coupled-model theory treatment of FBG\cite{Erdogan,FBG_decay,Grattan} to narrow our feature selection down to only those features that are dependent on the grating refractive index. 
\par The refractive index of the grating can be written as\cite{FBG_decay}:
\begin{equation}
n = n_{\text{eff}}^o +\Delta n_{\text{\text{eff}}}^{\text{mean} }+ \Delta n_{\text{eff}}^{\text{mod}}cos(\frac{2\pi z}{\Lambda})
\end{equation}
where $n^o_{\text{eff}}$ is the effective index of the unperturbed fiber, $\Delta n_{\text{eff}}^{\text{mean}}$ and $\Delta n_{\text{eff}}^{\text{mod}}$ are the "DC" (period-average) and "AC" (sinusodal change over the period) components of the effective index, respectively\cite{FBG_decay,Erdogan}. The AC component can be evaluated by examining the maximum reflectivity ($\Delta n_{\text{eff}}^{\text{mod}} = \frac{\lambda_B \tanh^{-1}(\sqrt(R_{\text{max}})}{\pi l} $) or in the case of highly reflective gratings where reflectance does not appear to be a sensitive measure of $\Delta n_{\text{eff}}^{\text{mod}}$,  FWHM of the grating spectra can be used. The FWHM of the grating response is known to vary linearly with changes in index modulation. The DC component of the refractive index change of a grating is given by $\Delta n_{\text{eff}}^{\text{mean}} =  \frac{\Delta\lambda_B}{2\Lambda}$and can be measured by tracking the detunning of the grating center wavelength away from the designed wavelength\cite{FBG_decay,Erdogan,Erdogan_decay}.

We therefore restrict ourselves to core group of endogenous variables derived from spectral features- peak center, maximum reflectance, FWHM and kurtosis (stand-in for systematic variations in index along the grating length) to construct potential models. As a baseline model we use simple linear regression between $(\lambda_B - \lambda_B^{\text{(298 K)}})$ and temperature as a quadratic function (Table 1).\footnote{We used detuning as the independent parameter since it shows stronger correlation with temperature than peak center. See supplemental for details}  We designate the first ascending ramp as the calibration run treating it as our training data upon which the regression model is trained. The remaining data is used as "out-of-sample" validation set to not only evaluate how well the trained model generalizes the sensor response but also to characterize and quantify the impact of thermally induced hysteresis. The baseline model indicates an average training error of 513 mK and out-of sample error of 878 mK. Ten of the 14 sensors examined here show significant thermal hysteresis or ageing effects (training error= 461 mK and out-of-sample error = 1040 mK). In these sensors hysteresis appears to be additive resulting in an offset error that shifts the intercept indicating the overall $n_{\text{eff}}$ is increasing as the sensor is exposed to elevated temperature (Fig 1). 
\begin{figure}[htbp!]
\centering\includegraphics[width=12 cm]{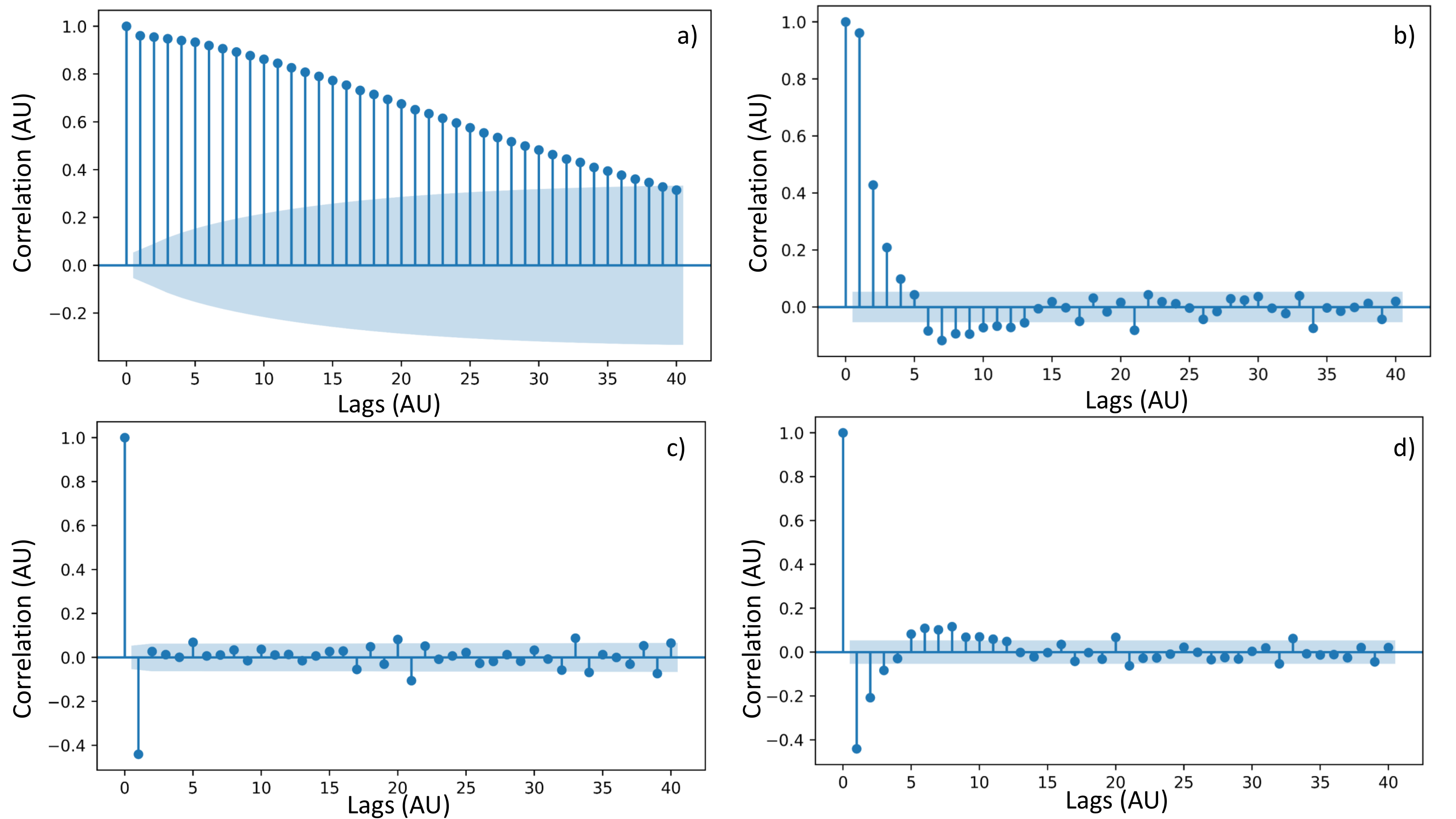}
\caption{ Autocorrelation function (ACF) of  S3 sensor's residual is plotted against computed time lags (a) and first-differenced residuals (c) are shown. b and d). Similarly on the right hand panel the partial-autocorrelation function (PACF) plot for S3 sensor's residual (b) and first-differenced residuals (d) are shown. The shaded region marks the uncertainty interval for the auto-correlation coefficients. These plots indicate that hysteresis in S3 is can be modeled as (1,1, [1,2,4,5]) process. See supplemental for ACF and PACF plots for every sensor examined in this study. }
\end{figure}
\par To improve upon our baseline model's performance we need to account for changes in $n_{\text{eff}}$ as the sensor is being thermally cycled. Equation 1 provides a framework for breaking down changes in $n_{\text{eff}}$ as arising from the AC or DC components of the grating. As noted above the aging effects result in redshift away from the grating wavelength at the start of thermal cycling, behaving as a DC refractive index change process. We model this redshift in baseline grating wavelength by employing a dynamic regression\cite{time_series_forecasting} model where the residuals from the training model are used to train a Autoregressive Integrated Moving Average (ARIMA)\cite{time_series_forecasting} whose output is added to training /calibration model's output.  We examined the auto correlation and partial auto-correlation plots of the residuals to determine a range of ($p,d,q$)\footnote{$p, d, q$ refer to the order of autoregressive, integrative and moving average processes. Hence a (1,1,1) process would be a described as being autoregressive in the first order with a simple linear trend and a  moving average of one. The first order difference i.e. linear trend correction is necessary to make the time series stationary} parameters for the ARIMA model that are needed to describe the time dependent behavior of the residuals (Fig 2). As shown in Fig 1d the time evolution of residuals is a non-stationary process as indicated by increasing trend. The presence of this trend is also captured by the ACF plot (Fig 2a) which shows positive values for the auto-correlation coefficient that slowly decrease as lags increase. In contrast, the first differenced residual time series' ACF plot is dominated by the first time lag indicating the time series can be transformed into a stationary process by taking the first difference. Furthermore, we note that a lack of significant peaks at longer time lags or multiples of time lags indicates a lack of cyclical behavior in the time evolution of the hysteresis.
\par The ACF and PACF plots of the difference residuals are used to determine the dominant terms for the autoregressive and moving average terms by selecting the highest time time lags correlation coefficients significantly larger than the calculated uncertainty. These parameters were then further optimized by maximizing the log-likelihood of the fit. Overall our exploration of parameter space indicates that simple ARIMA models containing a linear trend ($d=1$) and short term moving average ($q<10$) and autoregressive term ($p<2$) is sufficient to capture long term changes in the residual across all sensors. That is, the long-term drift in FBG can be described as consisting of three components: a slow linear drift, a short-term memory ($<$ 300 s) and a white noise component. Caution should be exercised when using the order of ARIMA terms to draw insights into physical processes responsible for the observed hysteresis. Given that changes in Bragg wavelength are due to changes in the $n_{eff}$, we interpret the linear trend as indicating the $n_{eff}$ of the sensor shows a slow, linear increase as thermal cycles progress. The origin(s) of the short-term memory effect is difficult to assign. It is likely that this term is capturing short term process such as modal noise due to thermal or strain relaxation in the fiber or short-term correlation in the temperature control loop of the bath.
\par As shown in Table 1 the mean uncertainty for one-step prediction is only 265 mK while uncertainty when using dynamic prediction\footnote{see supplemental for discussion of differences between one-step vs dynamic prediction} over the out-of-sample set is 623 mK which is 69\% and 41\% lower than the baseline model, respectively. We note that while the model performs well over the short-term (one-step forecast), uncertainty in the forecast grows with the horizon. As such these models are most appropriate over finite horizons.  For metrology problems requiring infinite horizons, state-space models or long-short term memory (LSTM) models\cite{lstm} that incorporate device physics, chemistry and thermal history may be more appropriate.
\par We evaluated the possibility that hysteresis maybe accounted by changes in the AC component i.e. the observed hysteresis may derive in part from the grating contrast erasure. As noted above, changes in the AC component of index proportionally impact the amplitude and width of the resonance spectra. In order to incorporate the AC index change into our model, we therefore incorporate additional features (time normalized amplitude, fractional FWHM and kurtosis)\footnote{kurtosis is a stand-in variable for non-uniform changes in AC index along the fiber axis i.e. it reports on the dephasing of the grating index contrast. time-normalized amplitude and fractional width are used in place of amplitude and FWHM, respectively, because they show stronger correlations with peak center.} into a multivariate regression model employing L2 regularization ($\alpha =  0.000001 $), multi-layer preceptrons (MLP) with one or two hidden layers employing sigmodial activation ([100x1]  or [5x1],[3x1] , respectively) and a hidden-state-like model where amplitude, fractional width and kurtosis are transformed and regressed to predict peak center detunning, $(\lambda_B - \lambda_B^{(298 K)})$, which in turn is added as a feature along with measured peak center detunning to a multivariate regression model. As shown in Table 2 the resulting models fail to improve upon the baseline model, generally performing worse. Failure of these models to accurately compensate for hysteresis suggests that while spectral features other than the peak center show temporal changes, these changes are neither linearly correlated with temperature nor the observed peak center. Hysteresis in FBG can be adequately modeled as a DC-only process. 

\begin{threeparttable}\footnotesize
\caption{Results of Dynamic Regression Model for Hysteresis  Compensation} \label{tab:tech_sum}
    \begin{tabular}{ cp{1.3cm}p{1.3cm}p{1.3cm}p{1.3cm}p{1.3cm}p{1.3cm}p{1.3cm}p{2.3cm}p{1.3cm}p{0.4cm} }
        \toprule
        \multicolumn{1}{c}{}                   \\
        Sensor& Hysteric& Training error& Out of sample error& One step error& Dynamic pred. error& Uncorrected error over dynamic range& \\ 
         
         \midrule\\
        	 S1	& Yes & 0.1369	& 2.106	& 0.32	& 0.79	& 1.9&\\
             S2	& Yes & 0.352	& 1.553	& 0.09	& 0.78	& 1.45&\\
			S3	& Yes & 0.4683	& 1.06	& 0.15	& 0.42	& 0.69&\\
			S4	& Yes & 0.774	& 0.88	& 0.34	& 0.48	& 0.81&\\
			S5	& Yes & 0.4576	& 0.761	& 0.17	& 0.42	& 0.598&\\
			S6	& Yes & 0.2485	& 0.749 & 0.25	& 0.6	& 0.67&\\
			S7	& Yes & 0.5381	& 0.808	& 0.34	& 0.55	& 2.35&\\
			S8	& Yes & 0.8184	& 1.269	& 0.63	& 1.05	& 1.08&\\
			S9	& Yes & 0.4565	& 0.81 & 0.15	& 0.73	& 0.54&\\
			S10	& Yes & 0.3622	& 0.4 & 0.21	& 0.41	& 0.41&\\

         \midrule
         S11 & No & 0.7529	& 0.6985 & 0.29	& 0.47	& 0.73&\\
         S12 & No	& 0.5255	& 0.5745	& 0.16	& 0.6	& 0.66& \\
		S13 \tnote{a}	& No	& 0.445	& 0.447	& 0.29	& 0.48	& 0.45& \\
		S13 \tnote{b}	& No	& 0.424	& 0.431	& 0.28	& 0.57	& 0.43& \\
		S14	& No	& 0.9336	& 0.6275	& 0.25	& 0.85	& 0.83& \\

         \midrule\\
          	Mean \tnote{c} &  &0.461 & 1.04 & 0.265 & 0.623 & 1.05&\\
          	Mean \tnote{d} &   & 0.513 & 0.878 & 0.261	& 0.613	& 0.9& \\
          \bottomrule
          \end{tabular}
          \begin{tablenotes}
          \small
          \item[a] input power  less than 10 microwatt
            \item[b] input power 2.5 mW 
            \item[c] mean of hysteretic sensors only
            \item[d] mean of all sensors
                  
        \end{tablenotes}
    \end{threeparttable}    
\clearpage
\begin{threeparttable}\footnotesize
\caption{Modeling Hysteresis using an Expanded set of Spectral Features as Inputs to the Calibration Model}\label{tab:tech_sum2}
    \begin{tabular}{p{1.5cm}cp{1.5cm}p{.4cm}}
        \toprule
        \multicolumn{1}{c}{}                   \\
        Model   & Training  Error (K) & Out of Sample Error (K) & 
        \\
         \midrule\\
        	 Baseline\tnote{a}	& 0.513	& 0.878	&\\
             Hidden-state-like\tnote{b}	&  0.627	& 6.66	&\\
			Lasso \tnote{c} & 1.13	& 6.31	&\\
			MLP\tnote{d}	& 10.42	& 22.98	&\\
			MLP \tnote{e}	& 7.88	& 21.14	&\\

          \bottomrule
          \end{tabular}
          \begin{tablenotes}
          \small
          \item[a] peak center detuning is the only input
          \item[b] fractional width, kurtosis and amplitude are used to predict wavelength detunning. this prediction is used as an input to a regression model along with measured wavelength detunning to infer temperature
          \item[c] input features include wavelength detunning, kurtosis, fractional width and area, alpha = 0.000001; 
          \item[d] hidden layers [5*3], alpha = 0.0001
            \item[e] hidden layer [100], alpha = 0.0001
                      
        \end{tablenotes}
    \end{threeparttable}    
\clearpage

\section{Summary}
Long term hysteresis or ageing effects in photonic thermometers \cite{Ahmed_NCSLI,Sergey_FBG,PTB_FBG,PhCC} represent a significant measurement science challenge to the adoption of photonic thermometry in-lieu of resistance thermometers. In this study we demonstrate that guided by device physics we can deploy proven statistical techniques to model the ageing effects and successfully reduce the measurement uncertainty by up to 70\%. Our work here serves as a motivation to develop first principles based thermo-optic coefficient models that can be co-depolyed with ARIMA models in Kalman filter like predictor-corrector algorithms to reduce measurement uncertainty and gain mechanistic understanding of processes driving long and short term drift in sensor characteristics.

\section{Acknowledgments}
Author thanks Dan Samarov, Gilad Kusne, Tobias Herman and Tyrus Berry for helpful discussion. This work was funded by NIST-on-a-chip (NOAC) initiative. 

\section{Disclosures}
The author declares no conflicts of interest.

\medskip


\medskip
 \bibliographystyle{elsarticle-num} 
 \bibliography{cas-refs}

\begin{thebibliography}{10}
\expandafter\ifx\csname url\endcsname\relax
  \def\url#1{\texttt{#1}}\fi
\expandafter\ifx\csname urlprefix\endcsname\relax\def\urlprefix{URL }\fi
\expandafter\ifx\csname href\endcsname\relax
  \def\href#1#2{#2} \def\path#1{#1}\fi

\bibitem{Xu:14}
H.~Xu, M.~Hafezi, J.~Fan, J.~M. Taylor, G.~F. Strouse, Z.~Ahmed,
  \href{http://opg.optica.org/oe/abstract.cfm?URI=oe-22-3-3098}{Ultra-sensitive
  chip-based photonic temperature sensor using ring resonator structures}, Opt.
  Express 22~(3) (2014) 3098--3104.
\newblock \href {https://doi.org/10.1364/OE.22.003098}
  {\path{doi:10.1364/OE.22.003098}}.
\newline\urlprefix\url{http://opg.optica.org/oe/abstract.cfm?URI=oe-22-3-3098}

\bibitem{Preston}
H.~Preston-Thomas, The international temperature scale of 1990 (its-90),
  Metrologia 27~(1) (1990) 3--10.

\bibitem{MPK2019}
{Consultative Committee for Thermometry under the auspices of the International
  Committee for Weights and Measures}, {\em mise en pratique for the definition
  of the kelvin in the SI},
  \url{https://www.bipm.org/en/publications/mises-en-pratique } (2019).

\bibitem{Mangum_2002}
B.~W. Mangum, G.~F. Strouse, W.~F. Guthrie, R.~Pello, M.~Stock, E.~Renaot,
  Y.~Hermier, G.~Bonnier, P.~Marcarino, K.~S. Gam, K.~H. Kang, Y.-G. Kim, J.~V.
  Nicholas, D.~R. White, T.~D. Dransfield, Y.~Duan, Y.~Qu, J.~Connolly, R.~L.
  Rusby, J.~Gray, G.~J. Sutton, D.~I. Head, K.~D. Hill, A.~Steele, K.~Nara,
  E.~Tegeler, U.~Noatsch, D.~Heyer, B.~Fellmuth, B.~Thiele-Krivoj, S.~Duris,
  A.~I. Pokhodun, N.~P. Moiseeva, A.~G. Ivanova, M.~J. de~Groot, J.~F.
  Dubbeldam, \href{https://doi.org/10.1088/0026-1394/39/2/7}{Summary of
  comparison of realizations of the {ITS}-90 over the range 83.8058 k to
  933.473 k: {CCT} key comparison {CCT}-k3}, Metrologia 39~(2) (2002) 179--205.
\newblock \href {https://doi.org/10.1088/0026-1394/39/2/7}
  {\path{doi:10.1088/0026-1394/39/2/7}}.
\newline\urlprefix\url{https://doi.org/10.1088/0026-1394/39/2/7}

\bibitem{AHMED_IMEKO}
Z.~Ahmed,
  \href{https://www.sciencedirect.com/science/article/pii/S2665917421002713}{Role
  of quantum technologies in reshaping the future of temperature metrology},
  Measurement: Sensors 18 (2021) 100308.
\newblock \href {https://doi.org/https://doi.org/10.1016/j.measen.2021.100308}
  {\path{doi:https://doi.org/10.1016/j.measen.2021.100308}}.
\newline\urlprefix\url{https://www.sciencedirect.com/science/article/pii/S2665917421002713}

\bibitem{Krenek_ring_resonator}
R.~Eisermann, S.~Krenek, G.~Winzer, S.~Rudtsch,
  \href{https://doi.org/10.1515/teme-2021-0054}{Photonic contact thermometry
  using silicon ring resonators and tuneable laser-based spectroscopy}, tm -
  Technisches Messen 88~(10) (2021) 640--654.
\newblock \href {https://doi.org/doi:10.1515/teme-2021-0054}
  {\path{doi:doi:10.1515/teme-2021-0054}}.
\newline\urlprefix\url{https://doi.org/10.1515/teme-2021-0054}

\bibitem{Sergey_ring}
S.~Janz, R.~Cheriton, D.-X. Xu, A.~Densmore, S.~Dedyulin, A.~Todd, J.~H.
  Schmid, P.~Cheben, M.~Vachon, M.~K. Dezfouli, D.~Melati,
  \href{http://opg.optica.org/oe/abstract.cfm?URI=oe-28-12-17409}{Photonic
  temperature and wavelength metrology by spectral pattern recognition}, Opt.
  Express 28~(12) (2020) 17409--17423.
\newblock \href {https://doi.org/10.1364/OE.394642}
  {\path{doi:10.1364/OE.394642}}.
\newline\urlprefix\url{http://opg.optica.org/oe/abstract.cfm?URI=oe-28-12-17409}

\bibitem{Yije_Fano}
C.~Zhang, G.~Kang, Y.~Xiong, T.~Xu, L.~Gu, X.~Gan, Y.~Pan, J.~Qu,
  \href{http://opg.optica.org/oe/abstract.cfm?URI=oe-28-9-12599}{Photonic
  thermometer with a sub-millikelvin resolution and broad temperature range by
  waveguide-microring fano resonance}, Opt. Express 28~(9) (2020) 12599--12608.
\newblock \href {https://doi.org/10.1364/OE.390966}
  {\path{doi:10.1364/OE.390966}}.
\newline\urlprefix\url{http://opg.optica.org/oe/abstract.cfm?URI=oe-28-9-12599}

\bibitem{Silicon_nitride}
C.~Zhang, G.-G. Kang, J.~Wang, S.~Wan, C.-H. Dong, Y.-J. Pan, J.-F. Qu,
  \href{https://www.sciencedirect.com/science/article/pii/S0263224121013774}{Photonic
  thermometer by silicon nitride microring resonator with milli-kelvin
  self-heating effect}, Measurement 188 (2022) 110494.
\newblock \href
  {https://doi.org/https://doi.org/10.1016/j.measurement.2021.110494}
  {\path{doi:https://doi.org/10.1016/j.measurement.2021.110494}}.
\newline\urlprefix\url{https://www.sciencedirect.com/science/article/pii/S0263224121013774}

\bibitem{Smith:21}
J.~A. Smith, P.~Hill, C.~Klitis, M.~Sorel, P.~A. Postigo, L.~Weituschat, M.~D.
  Dawson, M.~J. Strain,
  \href{http://opg.optica.org/abstract.cfm?URI=IPRSN-2021-IM1A.4}{High
  precision diamond-on-gan photonic thermometry enabled by transfer printing
  integration}, in: OSA Advanced Photonics Congress 2021, Optica Publishing
  Group, 2021, p. IM1A.4.
\newblock \href {https://doi.org/10.1364/IPRSN.2021.IM1A.4}
  {\path{doi:10.1364/IPRSN.2021.IM1A.4}}.
\newline\urlprefix\url{http://opg.optica.org/abstract.cfm?URI=IPRSN-2021-IM1A.4}

\bibitem{WBG_optics_letter}
N.~N. Klimov, S.~Mittal, M.~Berger, Z.~Ahmed,
  \href{http://opg.optica.org/ol/abstract.cfm?URI=ol-40-17-3934}{On-chip
  silicon waveguide bragg grating photonic temperature sensor}, Opt. Lett.
  40~(17) (2015) 3934--3936.
\newblock \href {https://doi.org/10.1364/OL.40.003934}
  {\path{doi:10.1364/OL.40.003934}}.
\newline\urlprefix\url{http://opg.optica.org/ol/abstract.cfm?URI=ol-40-17-3934}

\bibitem{PhCC}
N.~Klimov, T.~Purdy, Z.~Ahmed,
  \href{https://www.sciencedirect.com/science/article/pii/S0924424717318940}{Towards
  replacing resistance thermometry with photonic thermometry}, Sensors and
  Actuators A: Physical 269 (2018) 308--312.
\newblock \href {https://doi.org/https://doi.org/10.1016/j.sna.2017.11.055}
  {\path{doi:https://doi.org/10.1016/j.sna.2017.11.055}}.
\newline\urlprefix\url{https://www.sciencedirect.com/science/article/pii/S0924424717318940}

\bibitem{Ahmed_NCSLI}
Z.~Ahmed, J.~Filla, W.~Guthrie, J.~Quintavalle, Fiber bragg grating based
  thermometry, NCSLI Measure 10~(4) (2015) 28--31.
\newblock \href {https://doi.org/10.1080/19315775.2015.11721744}
  {\path{doi:10.1080/19315775.2015.11721744}}.

\bibitem{Sergey_FBG}
D.~Grobnic, C.~Hnatovsky, S.~Dedyulin, R.~B. Walker, H.~Ding, S.~J. Mihailov,
  \href{https://www.mdpi.com/1424-8220/21/4/1454}{Fiber bragg grating
  wavelength drift in long-term high temperature annealing}, Sensors 21~(4)
  (2021).
\newblock \href {https://doi.org/10.3390/s21041454}
  {\path{doi:10.3390/s21041454}}.
\newline\urlprefix\url{https://www.mdpi.com/1424-8220/21/4/1454}

\bibitem{PTB_FBG}
R.~Eisermann, S.~Krenek, T.~Habisreuther, P.~Ederer, S.~Simonsen, H.~Mathisen,
  T.~Elsmann, F.~Edler, D.~Schmid, A.~Lorenz, A.~A.~F. Olsen,
  \href{https://www.mdpi.com/1424-8220/22/3/1034}{Metrological characterization
  of a high-temperature hybrid sensor using thermal radiation and calibrated
  sapphire fiber bragg grating for process monitoring in harsh environments},
  Sensors 22~(3) (2022).
\newblock \href {https://doi.org/10.3390/s22031034}
  {\path{doi:10.3390/s22031034}}.
\newline\urlprefix\url{https://www.mdpi.com/1424-8220/22/3/1034}

\bibitem{Erdogan_decay}
T.~Erdogan, V.~Mizrahi, P.~J. Lemaire, D.~Monroe,
  \href{https://doi.org/10.1063/1.357062}{Decay of ultraviolet‐induced fiber
  bragg gratings}, Journal of Applied Physics 76~(1) (1994) 73--80.
\newblock \href {http://arxiv.org/abs/https://doi.org/10.1063/1.357062}
  {\path{arXiv:https://doi.org/10.1063/1.357062}}, \href
  {https://doi.org/10.1063/1.357062} {\path{doi:10.1063/1.357062}}.
\newline\urlprefix\url{https://doi.org/10.1063/1.357062}

\bibitem{FBG_decay}
H.~Patrick, S.~L. Gilbert, A.~Lidgard, M.~D. Gallagher,
  \href{https://doi.org/10.1063/1.360753}{Annealing of bragg gratings in
  hydrogen‐loaded optical fiber}, Journal of Applied Physics 78~(5) (1995)
  2940--2945.
\newblock \href {http://arxiv.org/abs/https://doi.org/10.1063/1.360753}
  {\path{arXiv:https://doi.org/10.1063/1.360753}}, \href
  {https://doi.org/10.1063/1.360753} {\path{doi:10.1063/1.360753}}.
\newline\urlprefix\url{https://doi.org/10.1063/1.360753}

\bibitem{FBGthesis}
D.~P. Hawn, The effects of high temperature and nuclear radiation on the
  optical transmission of silica optical fibers, Ph.D. thesis, The Ohio State
  University (2012).

\bibitem{Munko_2016}
A.~Munko, S.~Varzhel', S.~Arkhipov, A.~Gribaev, K.~Konnov, M.~Belikin,
  \href{https://doi.org/10.1088/1742-6596/735/1/012015}{The study of the
  thermal annealing of the bragg gratings induced in the hydrogenated
  birefringent optical fiber with an elliptical stress cladding}, Journal of
  Physics: Conference Series 735 (2016) 012015.
\newblock \href {https://doi.org/10.1088/1742-6596/735/1/012015}
  {\path{doi:10.1088/1742-6596/735/1/012015}}.
\newline\urlprefix\url{https://doi.org/10.1088/1742-6596/735/1/012015}

\bibitem{Grattan}
S.~Pal, J.~Mandal, T.~Sun, K.~T.~V. Grattan,
  \href{http://opg.optica.org/ao/abstract.cfm?URI=ao-42-12-2188}{Analysis of
  thermal decay and prediction of operational lifetime for a type i
  boron-germanium codoped fiber bragg grating}, Appl. Opt. 42~(12) (2003)
  2188--2197.
\newblock \href {https://doi.org/10.1364/AO.42.002188}
  {\path{doi:10.1364/AO.42.002188}}.
\newline\urlprefix\url{http://opg.optica.org/ao/abstract.cfm?URI=ao-42-12-2188}

\bibitem{python3}
G.~Van~Rossum, F.~L. Drake, Python 3 Reference Manual, CreateSpace, Scotts
  Valley, CA, 2009.

\bibitem{pandas}
{W}es {M}c{K}inney, {D}ata {S}tructures for {S}tatistical {C}omputing in
  {P}ython, in: {S}t\'efan van~der {W}alt, {J}arrod {M}illman (Eds.),
  {P}roceedings of the 9th {P}ython in {S}cience {C}onference, 2010, pp. 56 --
  61.
\newblock \href {https://doi.org/10.25080/Majora-92bf1922-00a}
  {\path{doi:10.25080/Majora-92bf1922-00a}}.

\bibitem{Seaborn}
M.~L. Waskom, \href{https://doi.org/10.21105/joss.03021}{seaborn: statistical
  data visualization}, Journal of Open Source Software 6~(60) (2021) 3021.
\newblock \href {https://doi.org/10.21105/joss.03021}
  {\path{doi:10.21105/joss.03021}}.
\newline\urlprefix\url{https://doi.org/10.21105/joss.03021}

\bibitem{matplotlib}
J.~D. Hunter, Matplotlib: A 2d graphics environment, Computing in Science \&
  Engineering 9~(3) (2007) 90--95.
\newblock \href {https://doi.org/10.1109/MCSE.2007.55}
  {\path{doi:10.1109/MCSE.2007.55}}.

\bibitem{scikit-learn}
F.~Pedregosa, G.~Varoquaux, A.~Gramfort, V.~Michel, B.~Thirion, O.~Grisel,
  M.~Blondel, P.~Prettenhofer, R.~Weiss, V.~Dubourg, J.~Vanderplas, A.~Passos,
  D.~Cournapeau, M.~Brucher, M.~Perrot, E.~Duchesnay, Scikit-learn: Machine
  learning in {P}ython, Journal of Machine Learning Research 12 (2011)
  2825--2830.

\bibitem{statsmodels}
S.~Seabold, J.~Perktold, statsmodels: Econometric and statistical modeling with
  python, in: 9th Python in Science Conference, 2010, pp. 92--96.

\bibitem{Erdogan}
T.~Erdogan, Fiber grating spectra, Journal of Lightwave Technology 15~(8)
  (1997) 1277--1294.
\newblock \href {https://doi.org/10.1109/50.618322}
  {\path{doi:10.1109/50.618322}}.

\bibitem{time_series_forecasting}
R.~J. Hyndman, G.~Athanasopoulos, Forecasting:Principles and Practice, 2nd
  edition, OTexts, Melbourne, Australia, 2018.

\bibitem{lstm}
S.~Hochreiter, J.~Schmidhuber,
  \href{https://doi.org/10.1162/neco.1997.9.8.1735}{{Long Short-Term Memory}},
  Neural Computation 9~(8) (1997) 1735--1780.
\newblock \href
  {http://arxiv.org/abs/https://direct.mit.edu/neco/article-pdf/9/8/1735/813796/neco.1997.9.8.1735.pdf}
  {\path{arXiv:https://direct.mit.edu/neco/article-pdf/9/8/1735/813796/neco.1997.9.8.1735.pdf}},
  \href {https://doi.org/10.1162/neco.1997.9.8.1735}
  {\path{doi:10.1162/neco.1997.9.8.1735}}.
\newline\urlprefix\url{https://doi.org/10.1162/neco.1997.9.8.1735}

\end{thebibliography}


\begin{thebibliography}{1}
\expandafter\ifx\csname url\endcsname\relax
  \def\url#1{\texttt{#1}}\fi
\expandafter\ifx\csname urlprefix\endcsname\relax\def\urlprefix{URL }\fi
\expandafter\ifx\csname href\endcsname\relax
  \def\href#1#2{#2} \def\path#1{#1}\fi

\bibitem{Erdogan_decay}
T.~Erdogan, V.~Mizrahi, P.~J. Lemaire, D.~Monroe,
  \href{https://doi.org/10.1063/1.357062}{Decay of ultraviolet‐induced fiber
  bragg gratings}, Journal of Applied Physics 76~(1) (1994) 73--80.
\newblock \href {http://arxiv.org/abs/https://doi.org/10.1063/1.357062}
  {\path{arXiv:https://doi.org/10.1063/1.357062}}, \href
  {http://dx.doi.org/10.1063/1.357062} {\path{doi:10.1063/1.357062}}.
\newline\urlprefix\url{https://doi.org/10.1063/1.357062}

\bibitem{FBG_decay}
H.~Patrick, S.~L. Gilbert, A.~Lidgard, M.~D. Gallagher,
  \href{https://doi.org/10.1063/1.360753}{Annealing of bragg gratings in
  hydrogen‐loaded optical fiber}, Journal of Applied Physics 78~(5) (1995)
  2940--2945.
\newblock \href {http://arxiv.org/abs/https://doi.org/10.1063/1.360753}
  {\path{arXiv:https://doi.org/10.1063/1.360753}}, \href
  {http://dx.doi.org/10.1063/1.360753} {\path{doi:10.1063/1.360753}}.
\newline\urlprefix\url{https://doi.org/10.1063/1.360753}

\bibitem{islr}
G.~James, D.~Witten, T.~Hastie, R.~Tibshirani, An Introduction to Statistical
  Learning: with Applications in R., Springer, New York, USA, 2013.
\newblock \href {http://dx.doi.org/10.1007/978-1-4614-7138-7}
  {\path{doi:10.1007/978-1-4614-7138-7}}.

\bibitem{time_series_forecasting}
R.~J. Hyndman, G.~Athanasopoulos, Forecasting:Principles and Practice, 2nd
  edition, OTexts, Melbourne, Australia, 2018.

\bibitem{flockhart}
G.~M. Flockhart, R.~R. Maier, J.~S. Barton, W.~N. MacPherson, J.~D. Jones,
  K.~E. Chisholm, L.~Zhang, I.~Bennion, I.~Read, P.~D. Foote, Quadratic
  behavior of fiber bragg grating temperature coefficients, Applied optics
  43~(13) (2004) 2744--2751.

\bibitem{Ahmed_NCSLI}
Z.~Ahmed, J.~Filla, W.~Guthrie, J.~Quintavalle, Fiber bragg grating based
  thermometry, NCSLI Measure 10~(4) (2015) 28--31.
\newblock \href {http://dx.doi.org/10.1080/19315775.2015.11721744}
  {\path{doi:10.1080/19315775.2015.11721744}}.

\end{thebibliography}





\end{document}


\begin{frontmatter}
\title{Supplemental: Hysteresis Compensation in Temperature Response of Fiber Bragg Grating Thermometers Using Dynamic Regression}
\author[inst1]{Zeeshan Ahmed}
\affiliation[inst1]{organization={National Institute of Standards and Technology, Physical Measurement Laboratory, Sensor Science Division},            addressline={100 Bureau Drive}, 
            city={Gaithersburg},
            postcode={20899}, 
            state={MD},
            country={USA}}
           
\section{Methodology}
\subsection{Data Exploration}

The data exploration process began with examination of Pearson coefficient heatmaps (Fig SP1) and pairplots (Fig SP 2) of all continuous variables over the entire dataset and then parsed over each of the non-categorical variable. Furthermore, each feature’s temperature and time dependence was examined  separately to ascertain their behavior over time and temperature history. Data exploration indicates that against temperature the following feature show positive correlations: peak center, kurtosis, (fractional) peak width and grating type. No significant correlations were observed between temperature and the following features: time, amplitude, laser power and fringe visibility. Furthermore, we observe peak detuning (a feature created by subtracting the value of peak center value at 293.15 K at the start of measurements from all subsequent measurements) shows stronger correlation with temperature (0.91) as compared to peak center (0.77).

The temporal evolution of each feature during thermal cycling was examined for each sensor.  Thermal time history and attendant changes in peak center and fractional peak width for sensors S6 and S7 are shown in Fig SP 3 and SP 4, respectively.   We note that the time histories of the two exemplar sensors show significant differences in the magnitude of changes, suggesting the hysteresis evolution is sensitive to the sensor’s history which may  include manufacturing conditions, thermal annealing and other variables.  Furthermore, as shown in Fig  SP 3 and SP 4, changes in spectra-derived features are slow and cumulative, rising above the prevailing measurement noise only over long time periods.  As shown in Fig SP 5, the Allen deviation (ADEV) plots for sensor S5 and S7 show that the measurement uncertainty at any given temperature is dominated by 1/f  noise as evidenced by a linear decrease in variance with integration time. Based on these results we conclude that peak center drift due to hysteresis either occurs outside the observation times used in this study\footnote{note that we allow an equilibriation time of upto 20 mins once the temperature bath reaches the set temperature to eliminate any thermal gradients in the sensor. Hysteretic changes maybe be occurring during this deadtime or over time scales much longer than the mintues to hours long observation times used here} or that the impact of these changes on measurement variance is smaller than the impact of other processes over the observation times used for each temperature measurement. Based on these results and previous literature\cite{Erdogan_decay,FBG_decay} we conclude that the process responsible for hysteresis is likely similar to slow random diffusion of a polymer on a rough energy landscape marred by shallow traps. During thermal annealing the grating appears to be dephasing- as suggested by changes in peak center and width- which suggests that during annealing some of the shallow traps are wiped out and their population transferred to deeper traps. As the effective bandgap of the material redshifts, the refractive index of the optical fiber increases resulting in an offset error of the temperature calibration. In our subsequent analysis we therefore treat the impact of hysteresis as being a cumulative effect of the temperature ramp.

\subsection{Modelling Methodology}
The serial correlations or memory effects introduced by hysteresis (see discussion above), pose significant challenges to model evaluation. In any model training endeavor where the goal is to develop a deployable,  generalized model, it is important to appropriately balance the bias-variance tradeoff\cite{islr}. In sensor literature often the focus is on the training error e.g. the mean square error of fitting a function to the entire dataset is reported. Only optimizing the training error runs the risk of over-fitting the data creating a model that reduces residuals when fitted to the data it was trained on but fairs poorly when new data is introduced. To balance the tendency of models to over-learn, the data is divided between a training and validation set (sometimes alternatively referred to as testing set). The model is initially trained on the training set and subsequently evaluated on the validation set. The goal is to choose a model that minimizes both the training and validation scores i.e. a generalizable model that accounts for significant trends in the data without learning the noise. 
\par The presence of serial correlations in the data however make simple train-test split strategy- e.g. randomly selecting and setting aside $30\%$ of the data points for testing set- inappropriate. As shown in table 1, for 9 out of 14 sensors a train-test split results in out-of-sample errors (validation error) that are either similar or smaller than the training errors suggesting the noise is not appropriately balanced between the two datasets. Similar results are seen for k-fold cross validation (CV) where the data set is randomly broken into k sets. The k-fold CV approach has the benefit of minimizing the impact of outliers on the fit\cite{islr}. 
A common solution to the serial correlation problem is to implement time-series cross validation (TSCV)\cite{time_series_forecasting, islr}. Such an approach, however, makes it difficult to interperate the model in terms that are common in metrology e.g. hysteresis. We, therefore, implement a version of TSCV by dividing the data between the first up ramp (training set) and subsequent data (testing set). This ramp-by-ramp division of the data is appropriate because as shown above the effects of changes driving hysteresis are only significant over long time periods. Our approach divides the dataset in two long time periods, allowing us to isolate the uncertainty introduced by long term drift. The model evaluation on the training set is carried out using k-fold cross validation (k=5)\cite{islr}. Consistent with previous literature\cite{flockhart} we find that a quadratic function vastly outperforms a first order model, reducing MSE by $\approx75\%$. We note that cubic functions show a slight improvement ($13 \%$) over quadratic function. For the purposes of this study, however, we have chosen the simpler quadratic function to model the temperature-wavelength (peak center) relationship staying consistent with previous literature\cite{Ahmed_NCSLI, flockhart}. A detailed evaluation of different models where model complexity (number of free parameters) is weighed against its validation error is the subject of a forthcoming manuscript.        

\begin{threeparttable}\footnotesize
\caption{Training and out-of-sample (validation) error in train-test split evaluation of a first order regression model}\label{tab:tech_sum2}
    \begin{tabular}{p{1.5cm}cp{1.5cm}p{.4cm}}
        \toprule
        \multicolumn{1}{c}{}                   \\
        Sensor   & Training  Error (K) & Out of Sample Error (K) & 
        \\
         \midrule\\
        	S1	& 2.107	& 1.984	&\\
                S2	&  1.78 & 1.67 &\\
			S3  & 1.930	& 1.851 &\\
			S4 & 2.314	& 2.291 &\\
			S5 & 0.357	& 0.341	&\\
                S6	&0.994 &	1.023 	&\\
                S7	& 2.116&	2.099 	&\\
			S8  & 2.196&	2.057	&\\
			S9 & 2.301&	2.275	&\\
			S10 & 2.418&	2.472	&\\
                S11	& 2.330&	2.344	&\\
                S12	& 1.947&	1.937 &\\
			S13\tnote{a} & 2.138	&2.161	&\\
			S13\tnote{b}  & 2.322	&2.571	&\\
			S14 & 2.228 & 2.279 &\\

          \bottomrule
          \end{tabular}
          \begin{tablenotes}
          \small
          \item[a]1$\mu W$ input power
          \item[b] 2.5mW input power

        \end{tablenotes}
    \end{threeparttable}    
\clearpage
\subsubsection{Dynamic Regression}
As discussed in the manuscript we employed ARIMA models in a dynamic regression scheme to compensate for the long-term drift in the sensor refractive index due to hysteresis. As noted in the manuscript, the ($p,d,q$) terms for each sensor is determined by examining the ACF and PCF plots of the residuals and first-difference residuals \cite{time_series_forecasting}. The time dependent temperature changes, residuals and ACF and PCF plots for each sensor are shown below. We note that the ARIMA models are evaluated by computing their performance for one-step and dynamic prediction. The one-step prediction refers to outputs of the dynamic regression model where the validation set is introduced one-sequential time step at a time and prediction generated using the last time step i.e. predictions are restricted to short time-horizons. In the case of dynamic prediction, the entire validation set is evaluated at once i.e. the input to ARIMA model (residuals) are restricted to only the calibration/testing set. The results of one-step prediction represents the best-case performance that can be expected from these models. The dynamic prediction results, on the other hand, provides a measure of long-term (few weeks) performance of these models.

\begin{figure}[htbp!]
\centering\includegraphics[width=12 cm]{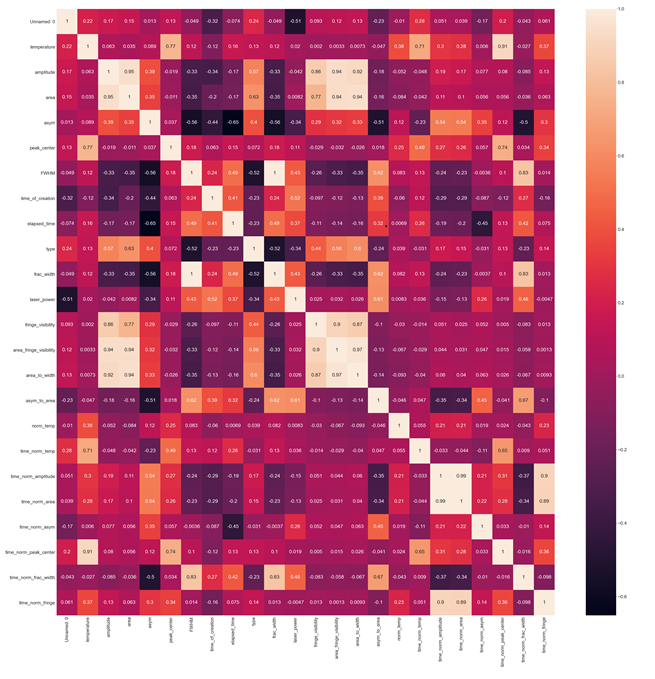}
\caption{Heatmap of Pearson correlation coefficient for non-categorical variables for the entire data of FBG temperature response}
\end{figure}

\begin{figure}[htbp!]
\centering\includegraphics[width=12 cm]{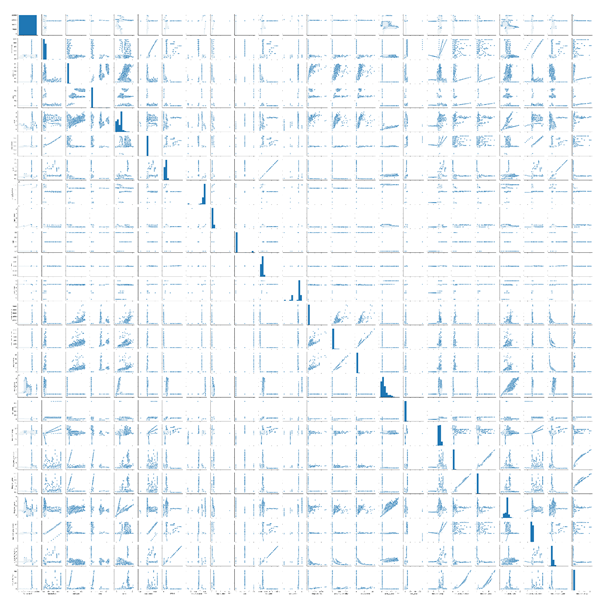}
\caption{Pairplots of all non-categorical variables over the entire data set}
\end{figure}

\begin{figure}[htbp!]
\centering\includegraphics[width=12 cm]{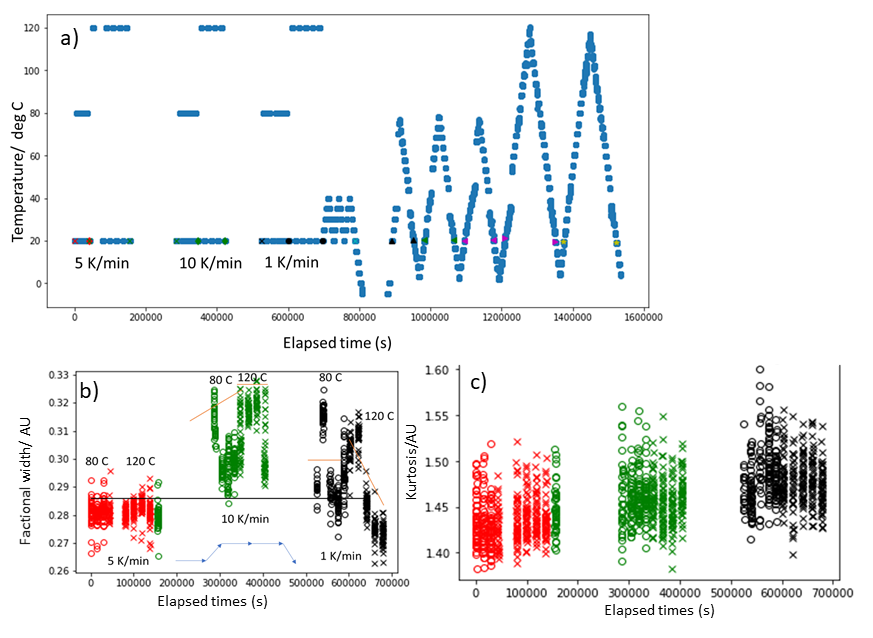}
\caption{a) Thermal cycling history of sensor S6 includes different heating rates and temperature excursions. At 293 K, time and temperature dependent evolution of fractional width b) and Kurtosis c) in sensor S6 shows significant memory effects }
\end{figure}

\begin{figure}[htbp!]
\centering\includegraphics[width=12 cm]{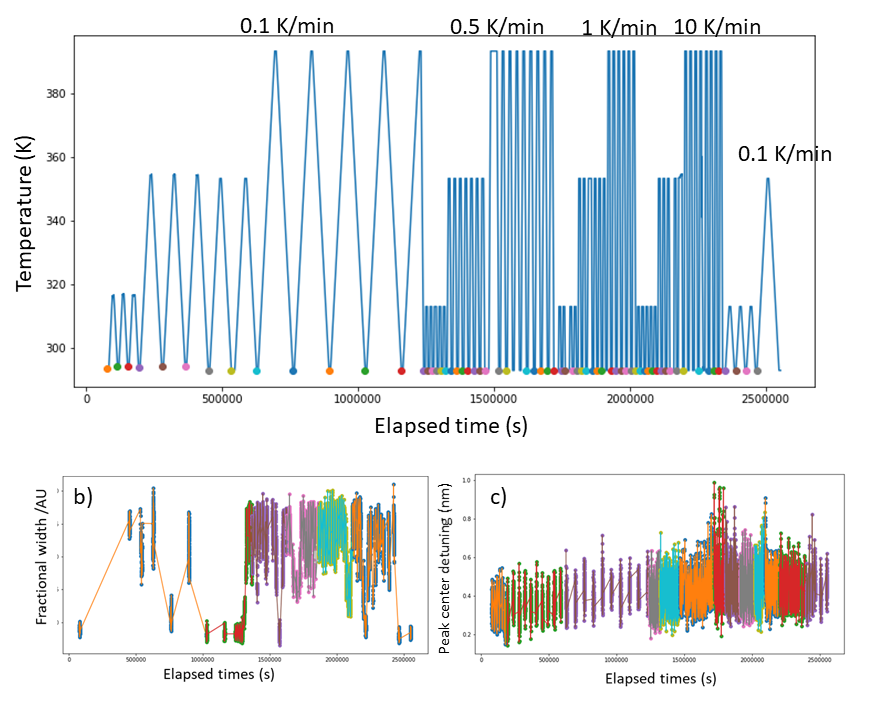}
\caption{a) Thermal cycling history of sensor S7 includes different heating rates and temperature excursions. At 293 K, the cumulative effective of thermal treatment's effect on the fractional width (a) and peak center detunning (b) can be seen.}
\end{figure}

\begin{figure}[htbp!]
\centering\includegraphics[width=12 cm]{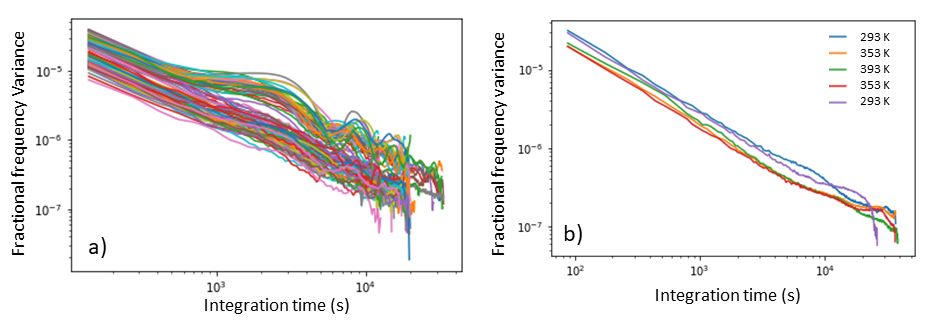}
\caption{a) Allen deviation plots for sensor S7 at every temperature between 283 K and 393 K recorded over three thermal cycles. Temperature was incremented in 5 K steps b) Allen deviation plots for sensor S5 at 293 K  (start), 353 K , 393 K  from the ramp-up and 293 K  (ramp-down) are shown. For both sensors, regardless of temperature, we observe a dominant linear trend indicating the measurement noise is dominated by 1/f noise process }
\end{figure}


\begin{figure}[htbp!]
\centering\includegraphics[width=12 cm]{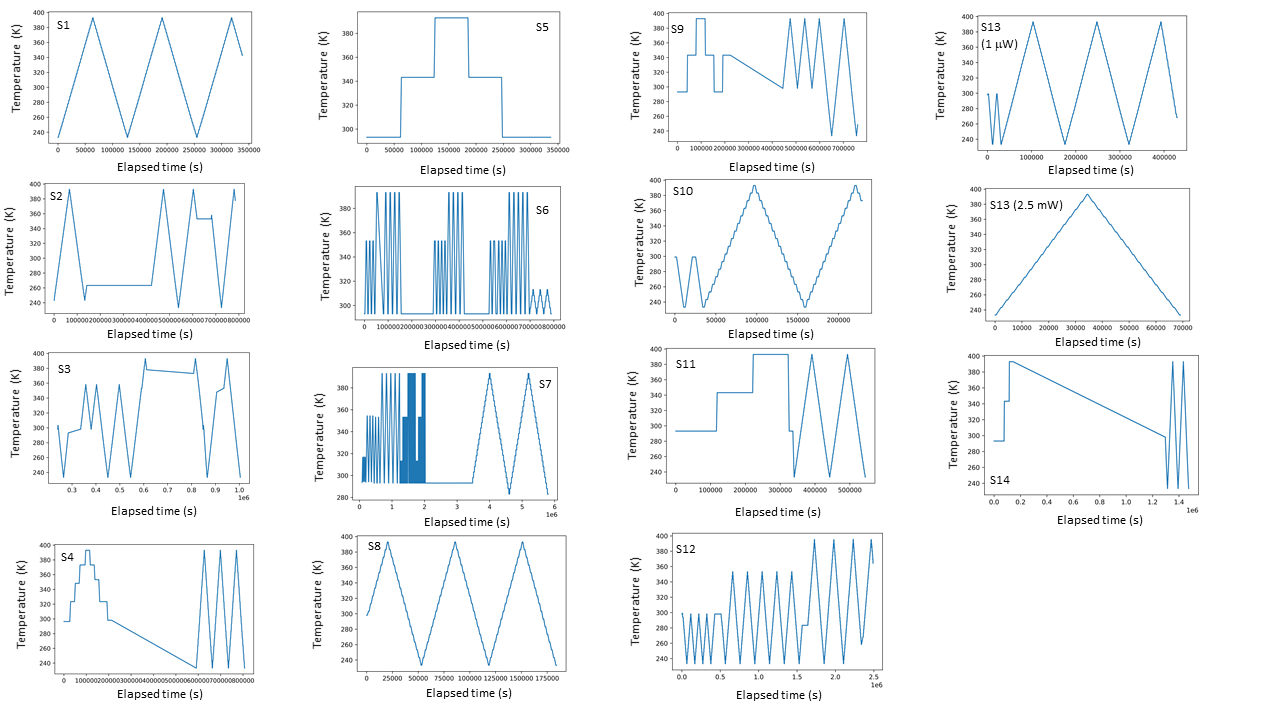}
\caption{Time history of temperature changes for each sensor is shown}
\end{figure}

\begin{figure}[htbp!]
\centering\includegraphics[width=12 cm]{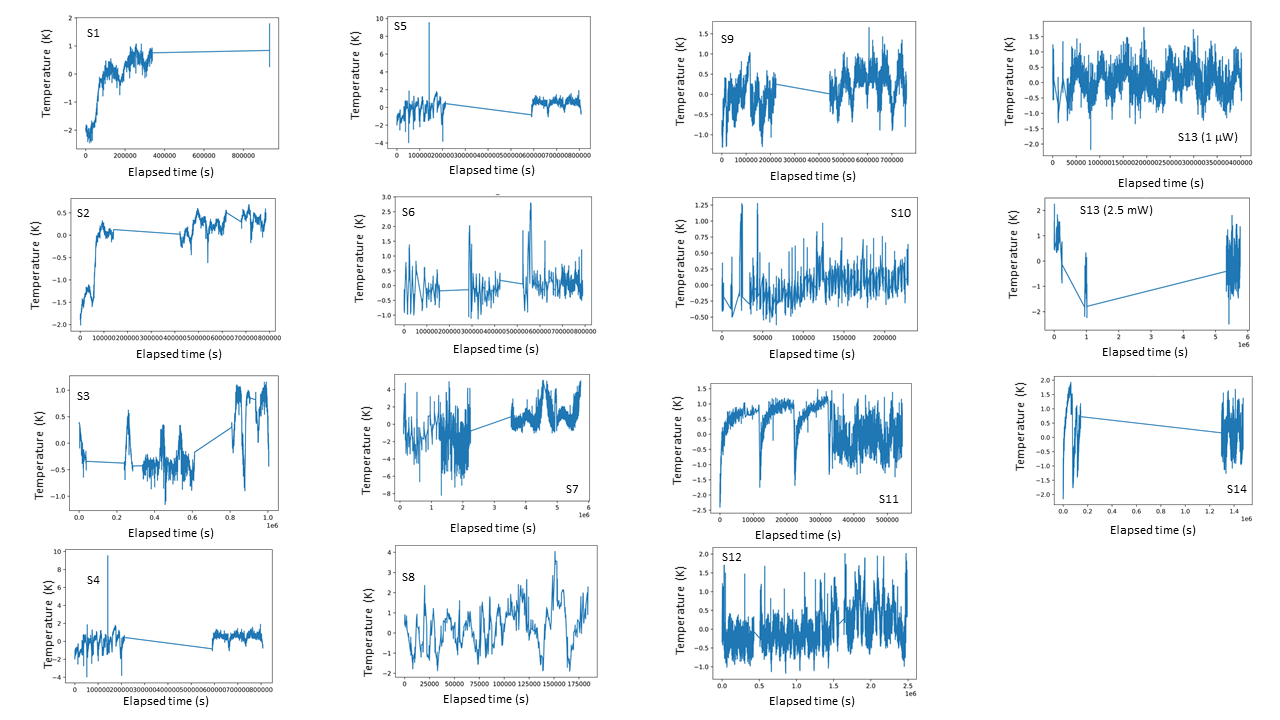}
\caption{Temporal evolution of residuals calculated from regression model for each sensor are shown}
\end{figure}

\begin{figure}[htbp!]
\centering\includegraphics[width=12 cm]{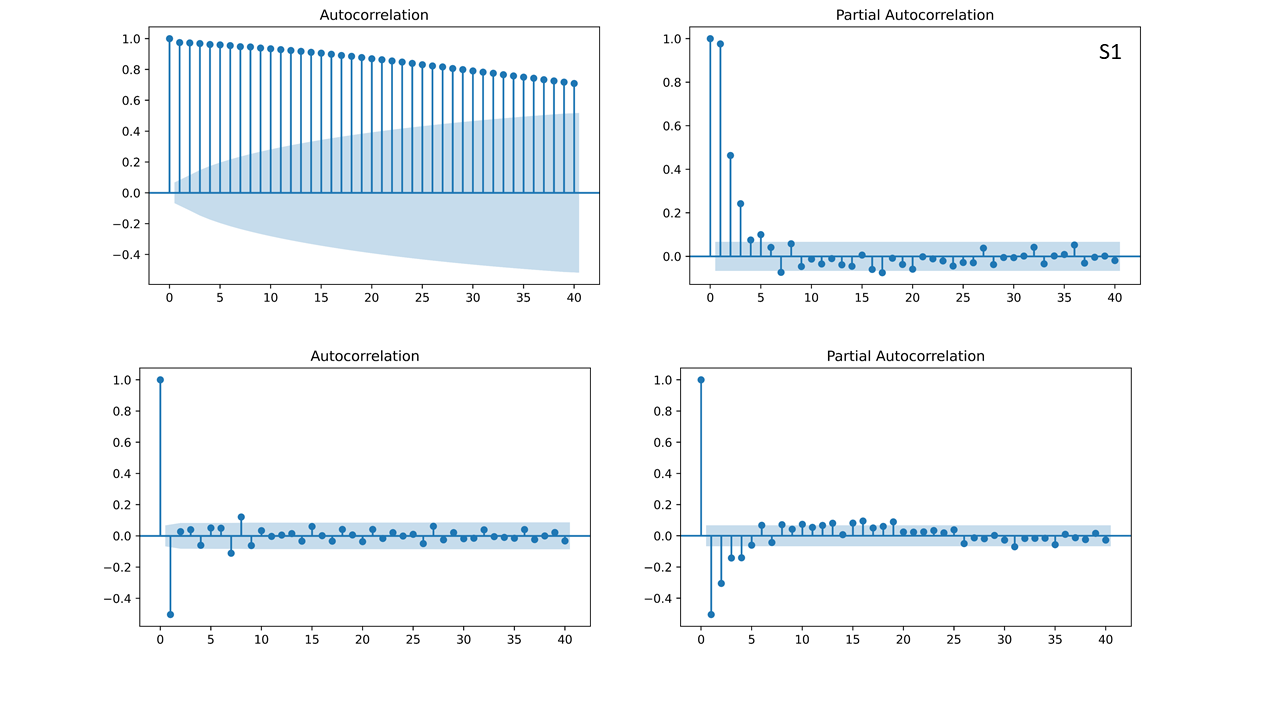}
\caption{ACF and PCF plots calculated form the residuals (top) and first-difference residuals plot (bottom) for S1 are shown.   }
\end{figure}

\begin{figure}[htbp!]
\centering\includegraphics[width=12 cm]{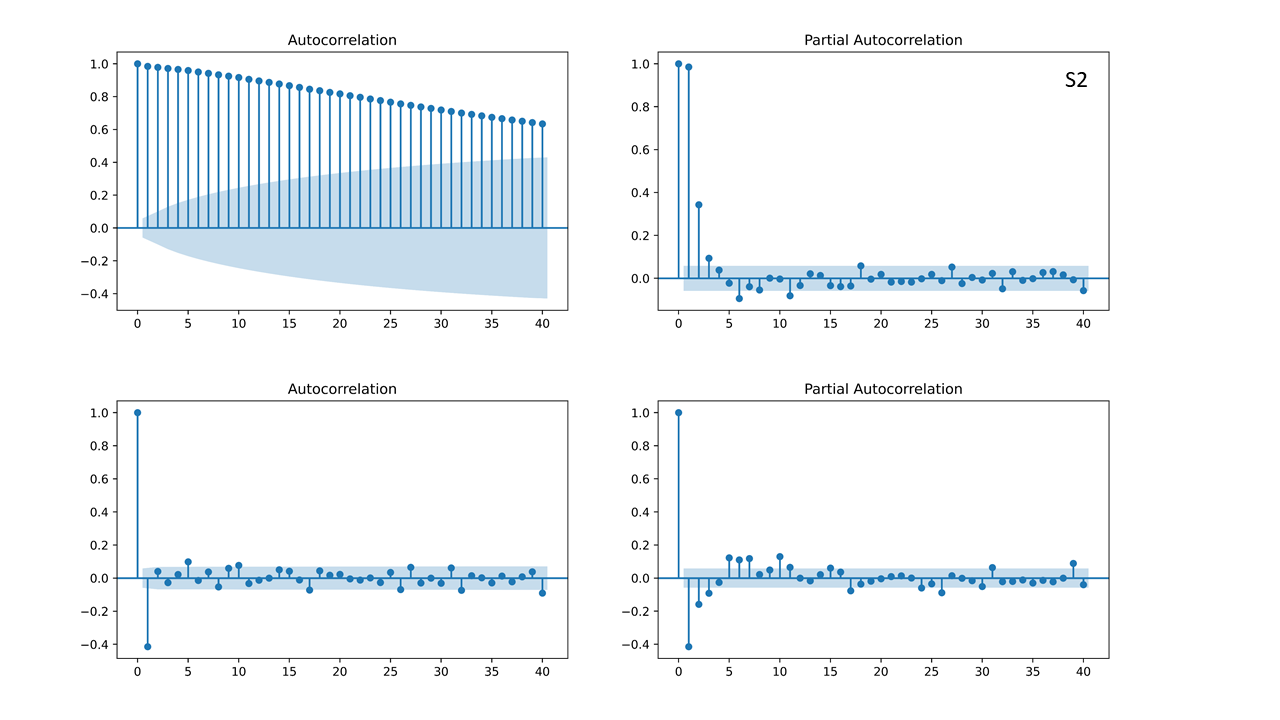}
\caption{ACF and PCF plots calculated form the residuals (top) and first-difference residuals plot (bottom) for S2 are shown.   }
\end{figure}

\begin{figure}[htbp!]
\centering\includegraphics[width=12 cm]{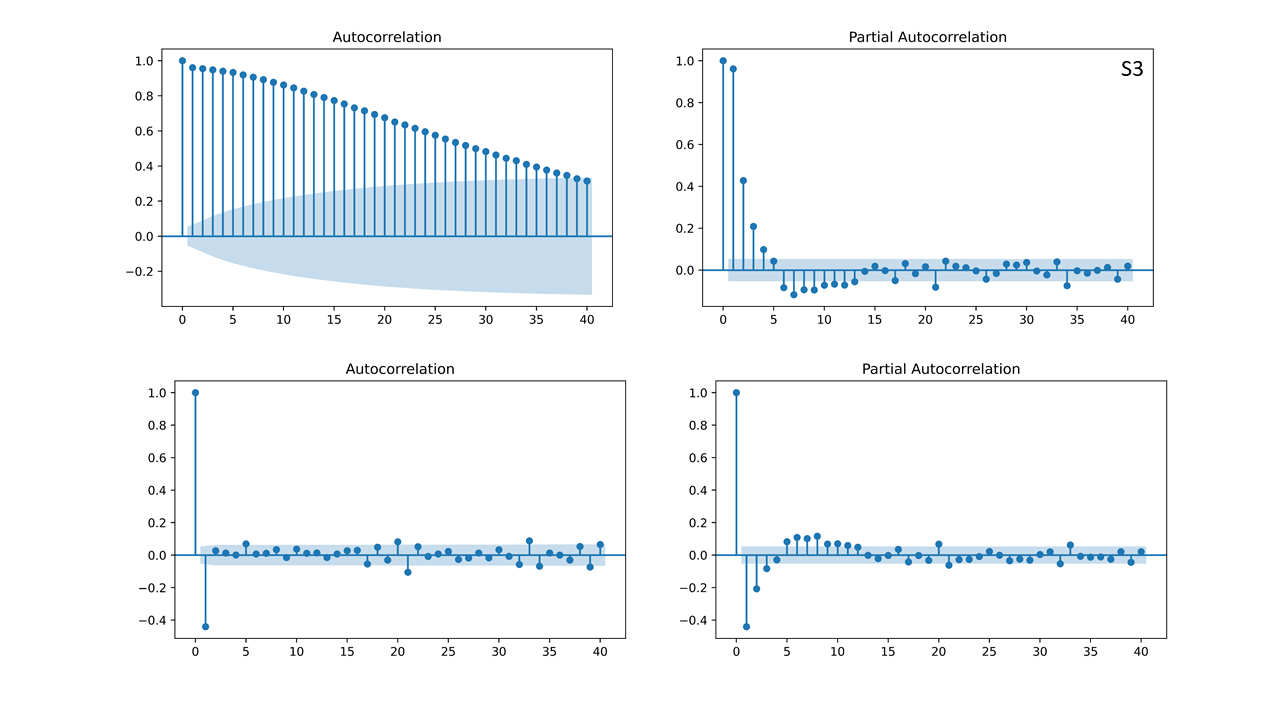}
\caption{ACF and PCF plots calculated form the residuals (top) and first-difference residuals plot (bottom) for S3 are shown.   }
\end{figure}

\begin{figure}[htbp!]
\centering\includegraphics[width=12 cm]{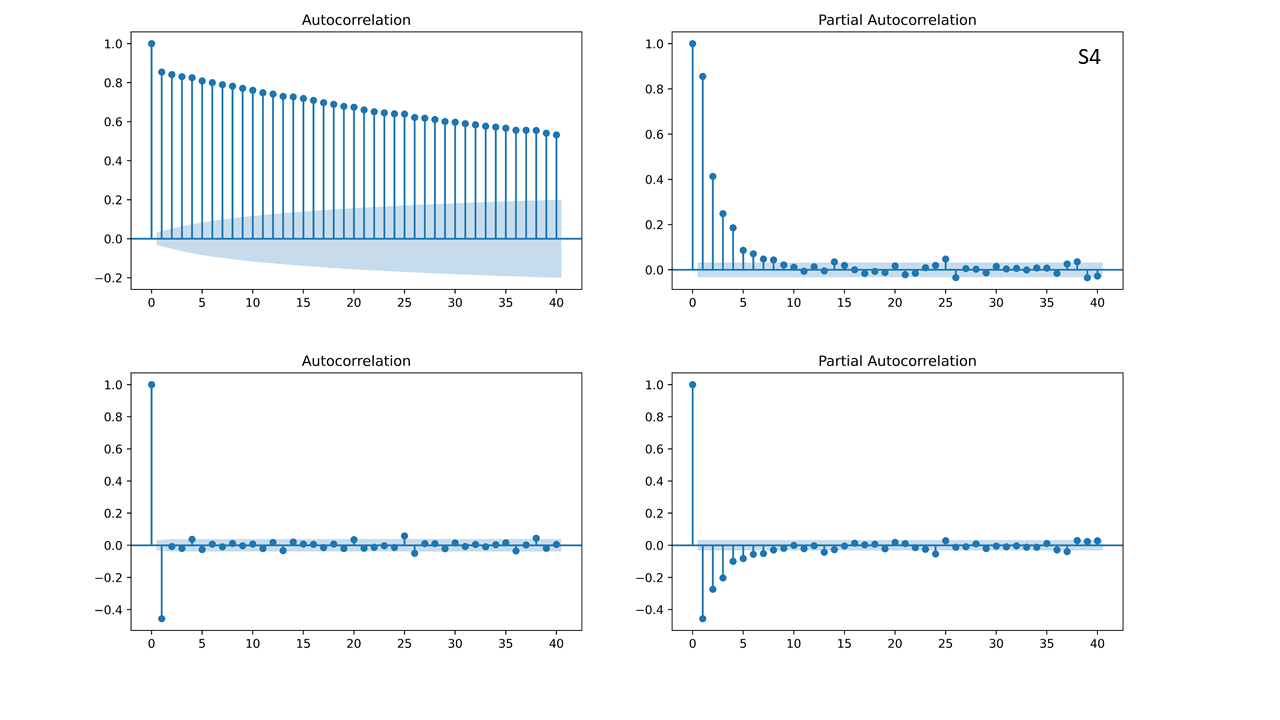}
\caption{ACF and PCF plots calculated form the residuals (top) and first-difference residuals plot (bottom) for S4 are shown.   }
\end{figure}

\begin{figure}[htbp!]
\centering\includegraphics[width=12 cm]{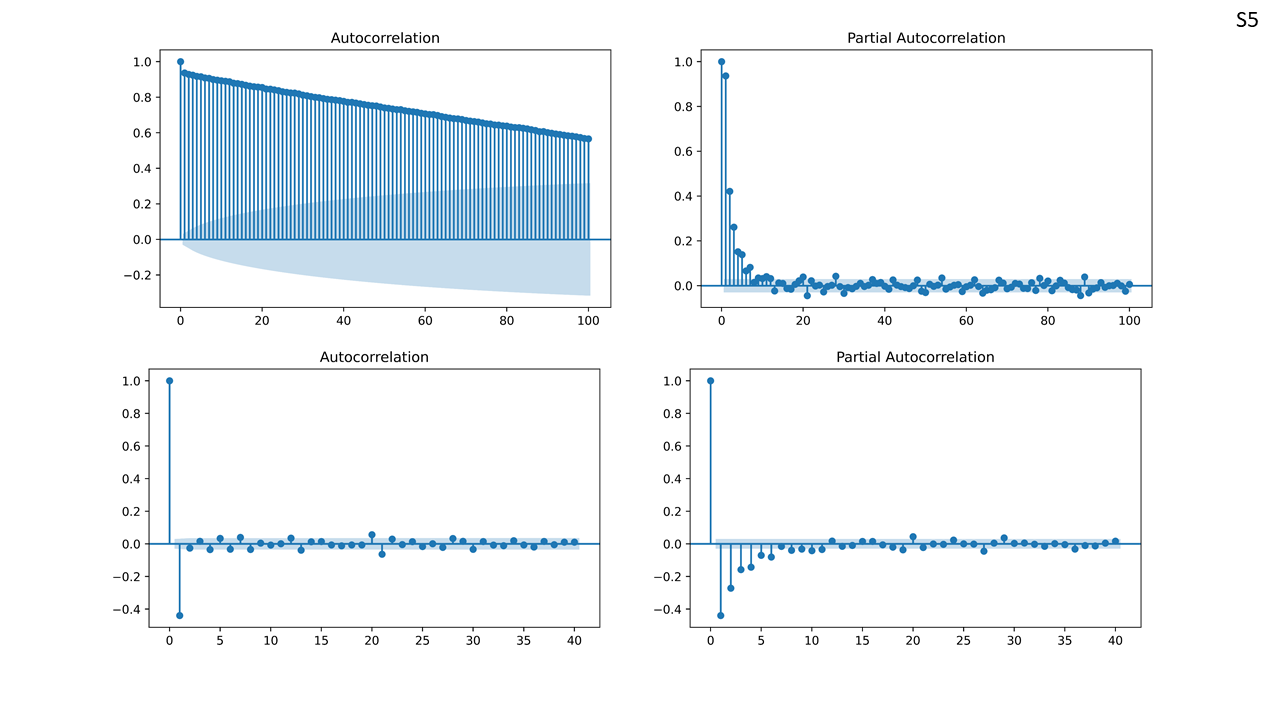}
\caption{ACF and PCF plots calculated form the residuals (top) and first-difference residuals plot (bottom) for S5 are shown.   }
\end{figure}

\begin{figure}[htbp!]
\centering\includegraphics[width=12 cm]{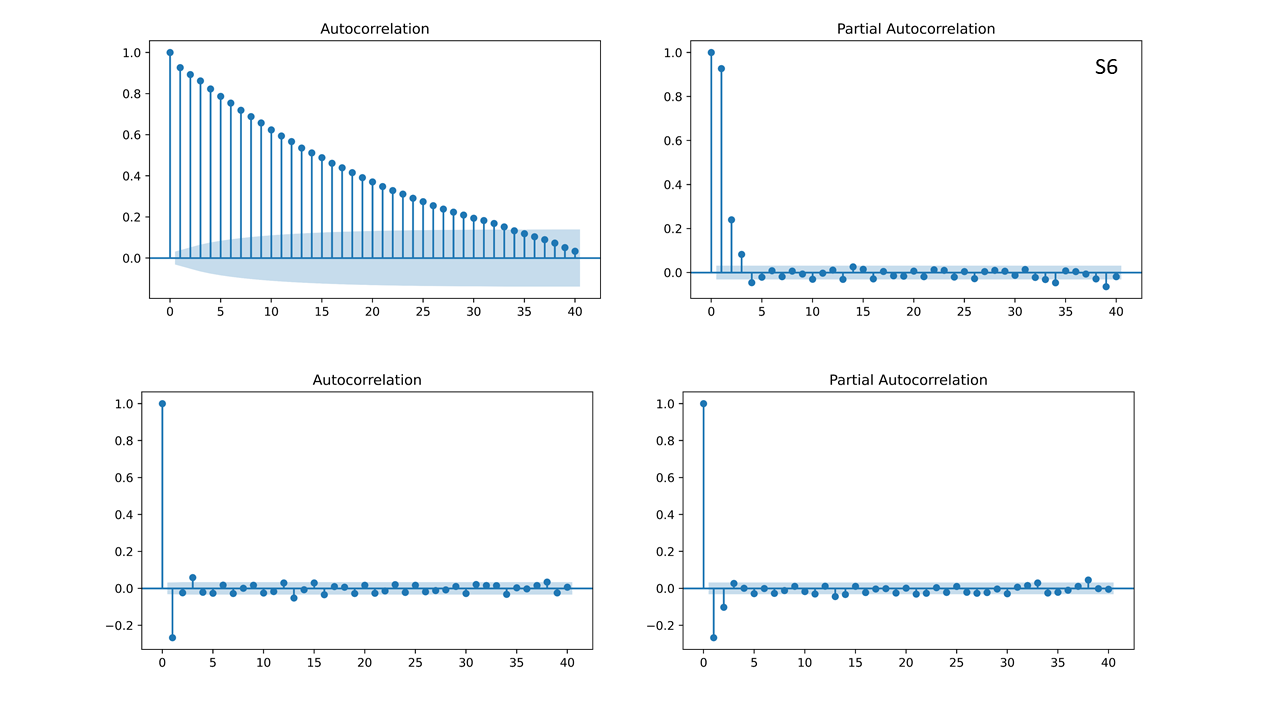}
\caption{ACF and PCF plots calculated form the residuals (top) and first-difference residuals plot (bottom) for S6 are shown.   }
\end{figure}

\begin{figure}[htbp!]
\centering\includegraphics[width=12 cm]{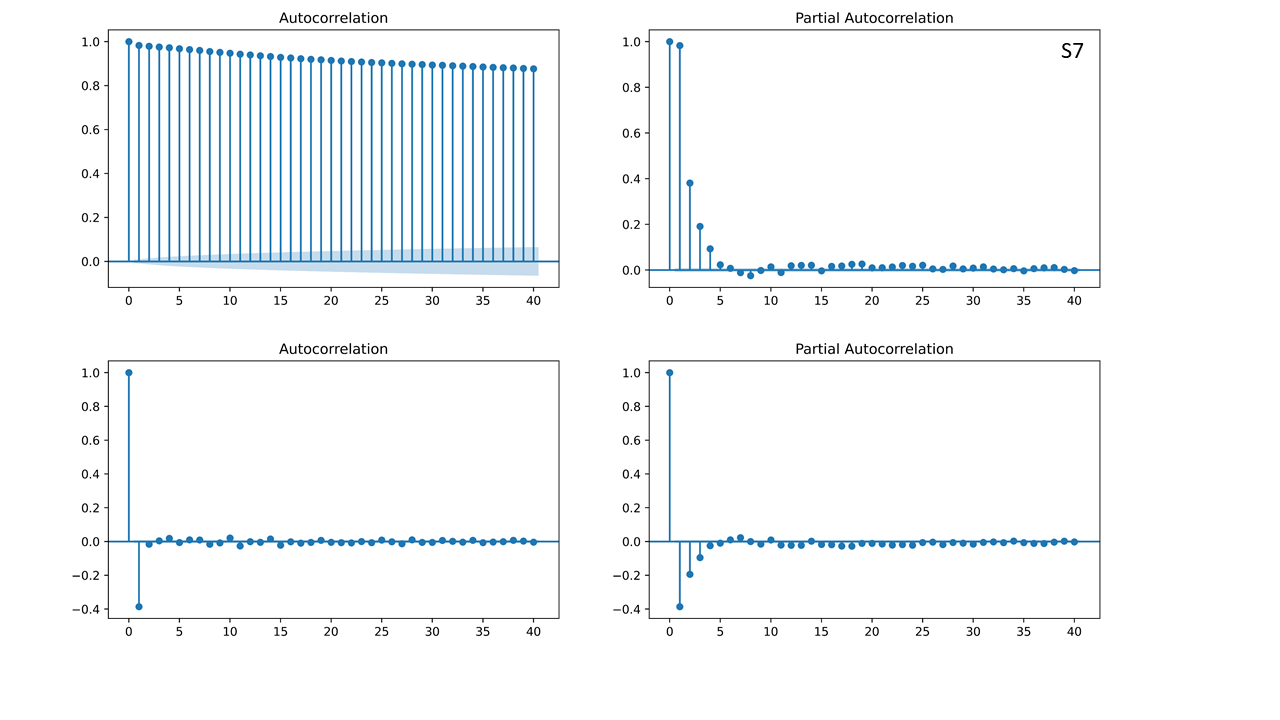}
\caption{ACF and PCF plots calculated form the residuals (top) and first-difference residuals plot (bottom) for S7 are shown.   }
\end{figure}

\begin{figure}[htbp!]
\centering\includegraphics[width=12 cm]{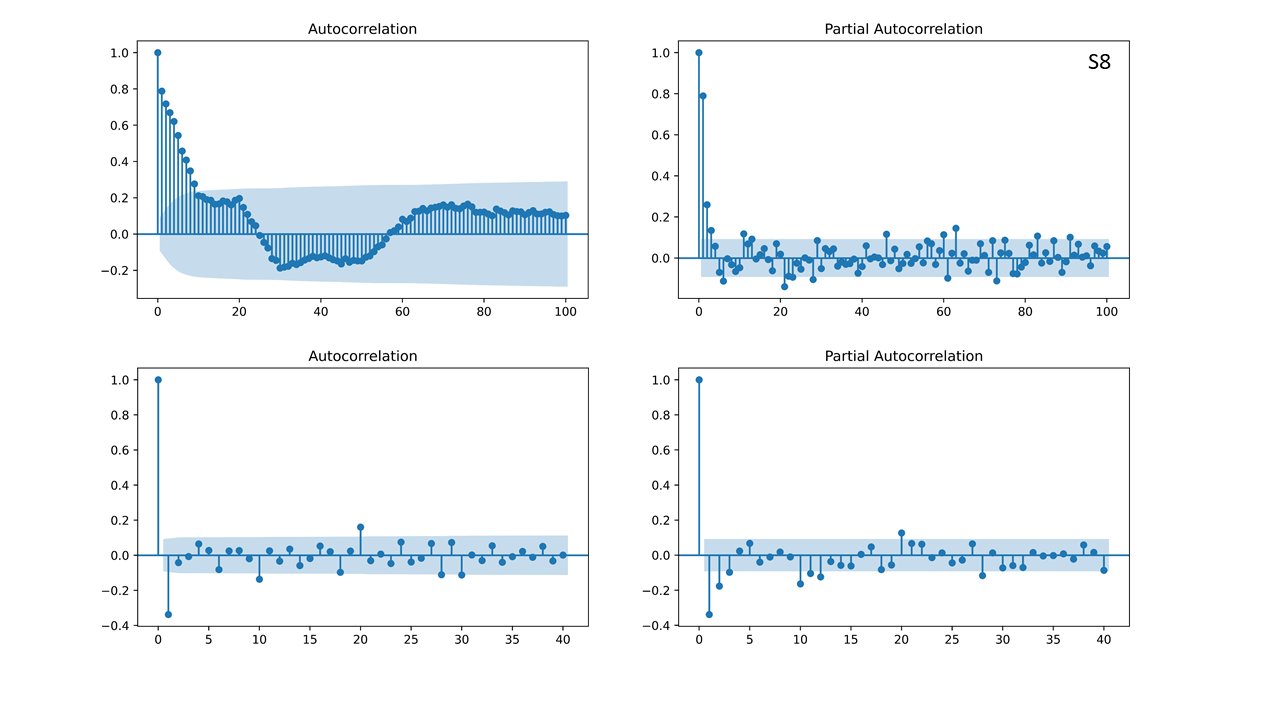}
\caption{ACF and PCF plots calculated form the residuals (top) and first-difference residuals plot (bottom) for S8 are shown.   }
\end{figure}

\begin{figure}[htbp!]
\centering\includegraphics[width=12 cm]{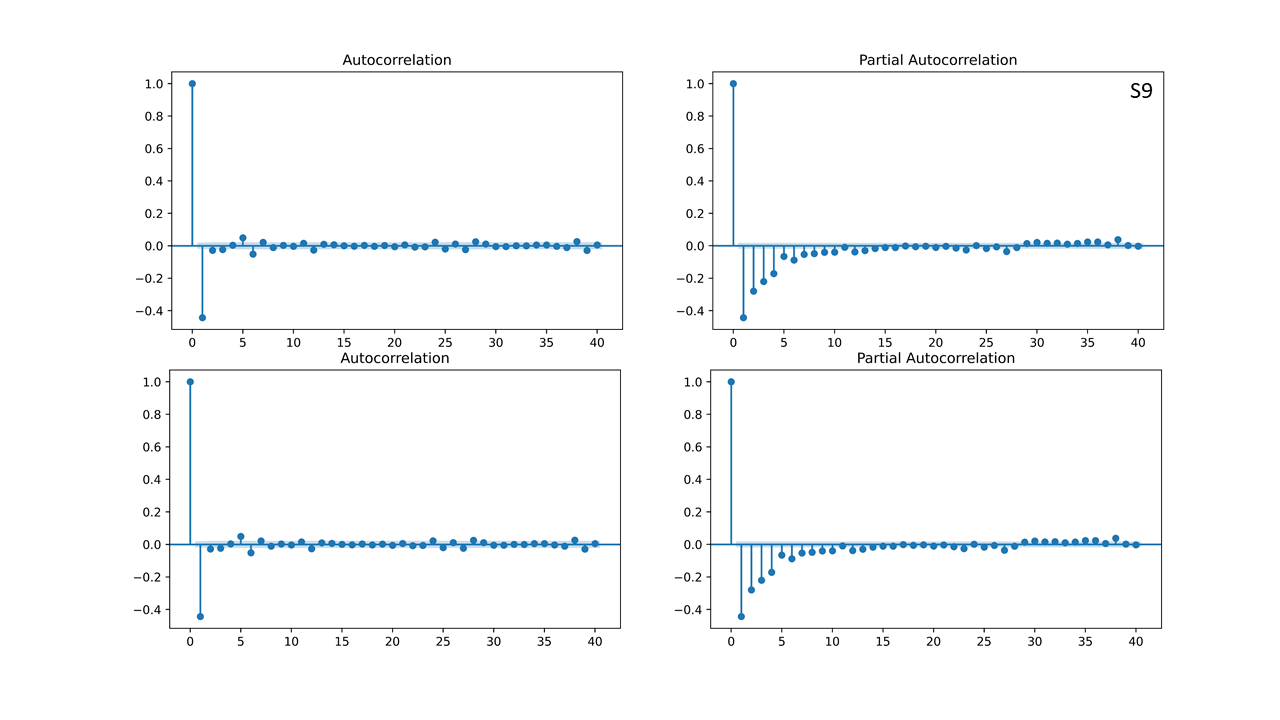}
\caption{ACF and PCF plots calculated form the residuals (top) and first-difference residuals plot (bottom) for S9 are shown.   }

\end{figure}\begin{figure}[htbp!]
\centering\includegraphics[width=12 cm]{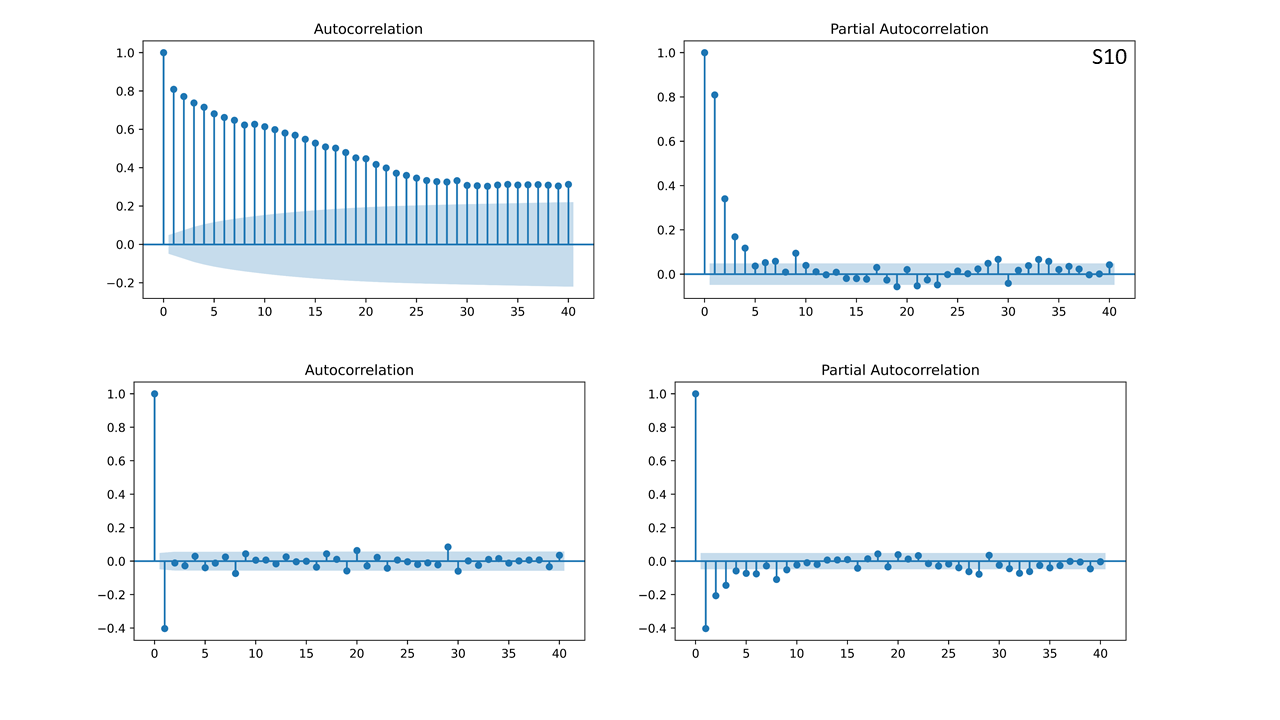}
\caption{ACF and PCF plots calculated form the residuals (top) and first-difference residuals plot (bottom) for S10 are shown.   }
\end{figure}

\begin{figure}[htbp!]
\centering\includegraphics[width=12 cm]{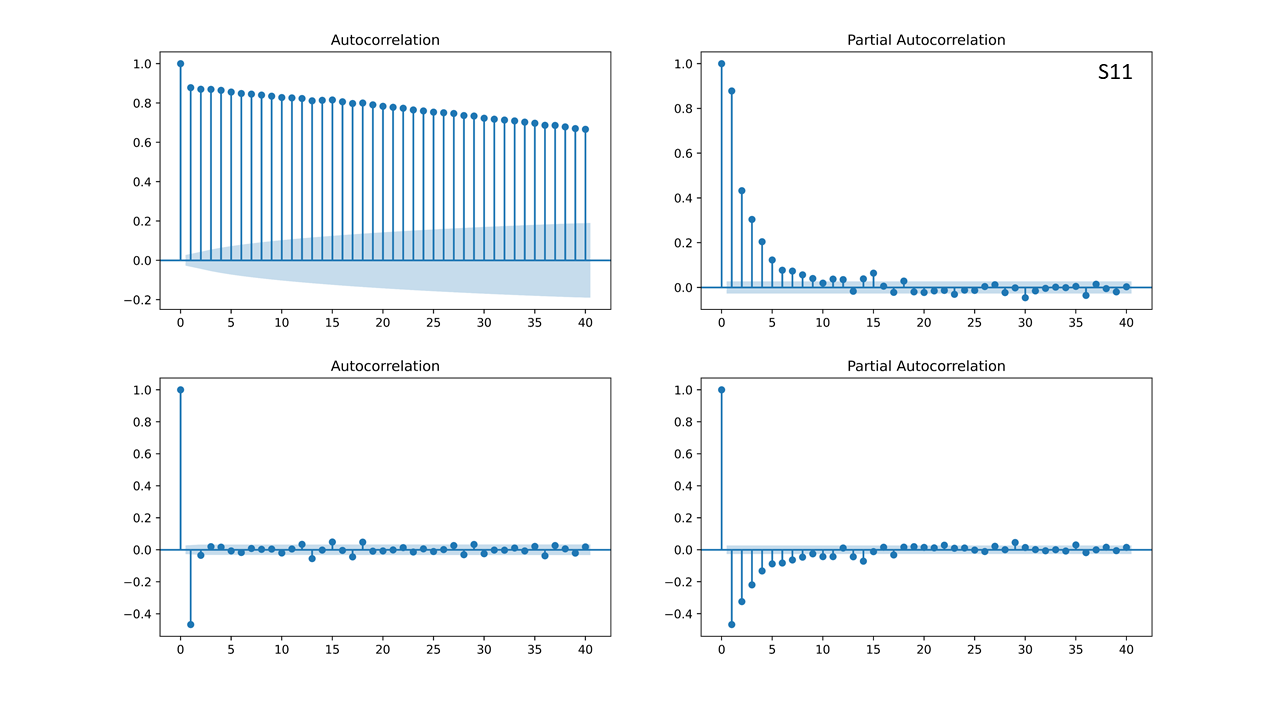}
\caption{ACF and PCF plots calculated form the residuals (top) and first-difference residuals plot (bottom) for S11 are shown.   }

\end{figure}\begin{figure}[htbp!]
\centering\includegraphics[width=12 cm]{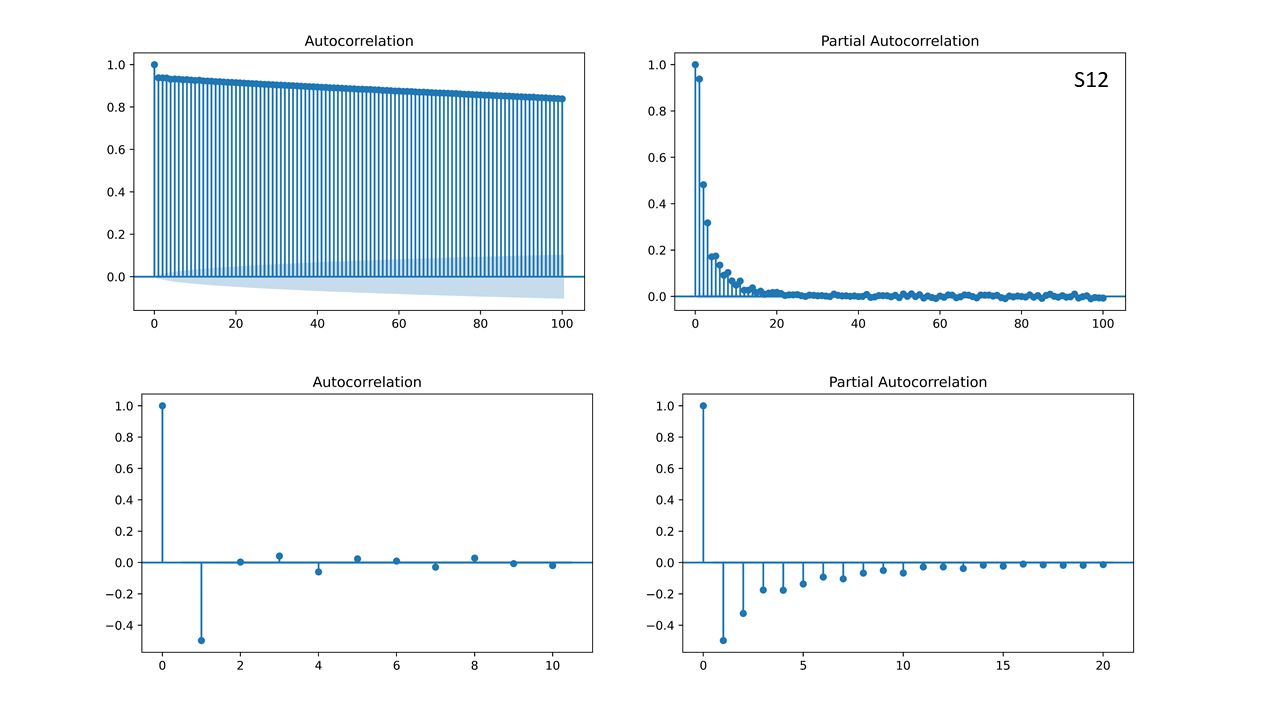}
\caption{ACF and PCF plots calculated form the residuals (top) and first-difference residuals plot (bottom) for S12 are shown.   }

\end{figure}\begin{figure}[htbp!]
\centering\includegraphics[width=12 cm]{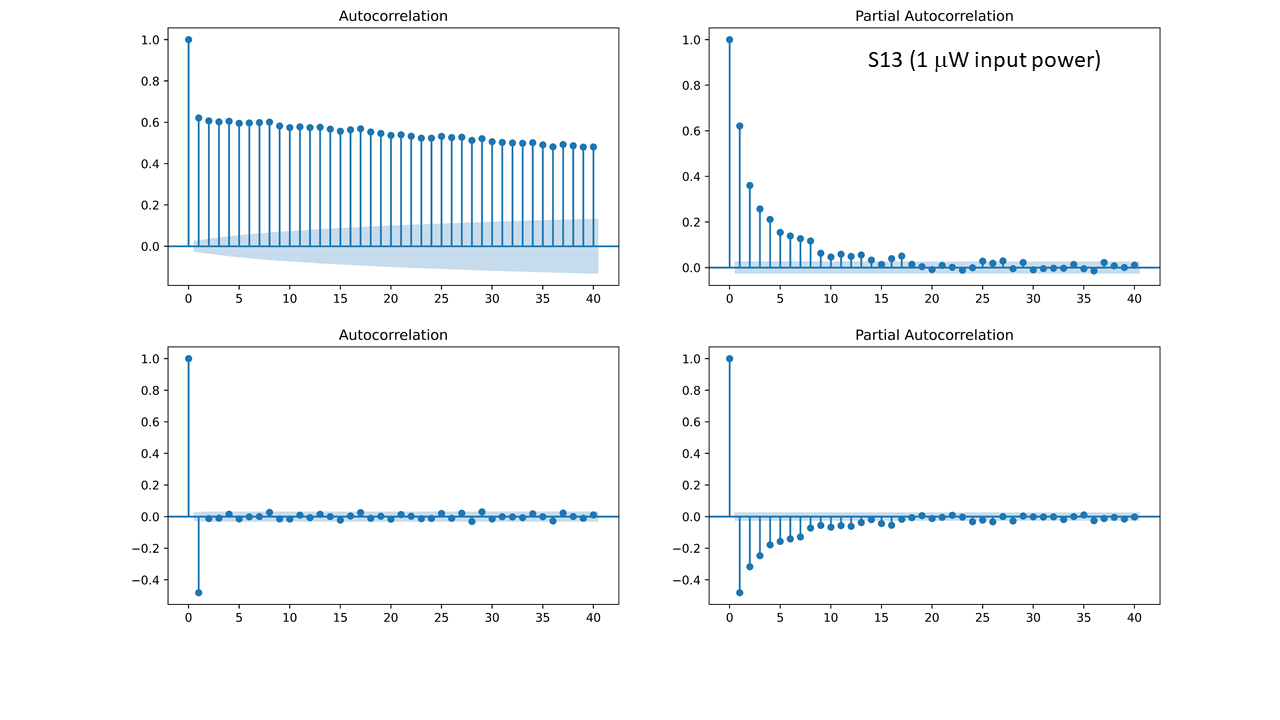}
\caption{ACF and PCF plots calculated form the residuals (top) and first-difference residuals plot (bottom) for S13 at 1$\mu W$ of input power are shown.   }

\end{figure}\begin{figure}[htbp!]
\centering\includegraphics[width=12 cm]{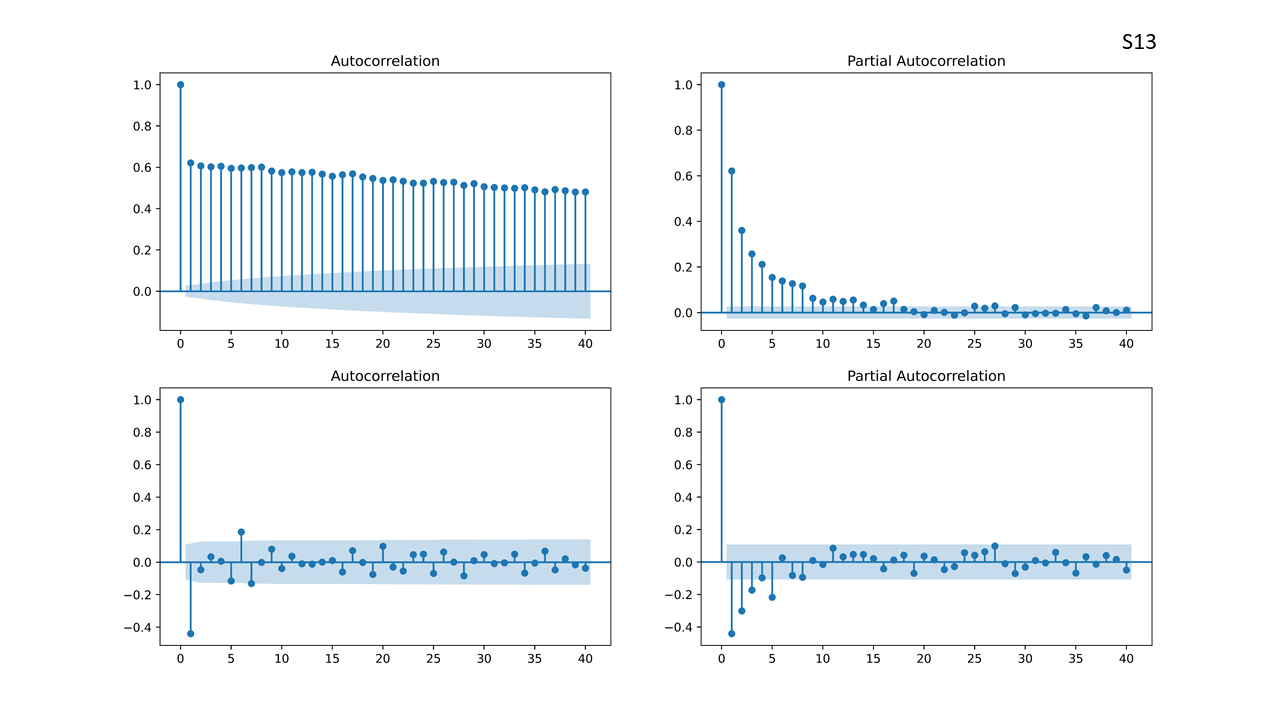}
\caption{ACF and PCF plots calculated form the residuals (top) and first-difference residuals plot (bottom) for S13 at 2.5 mW of input power are shown.   }

\end{figure}\begin{figure}[htbp!]
\centering\includegraphics[width=12 cm]{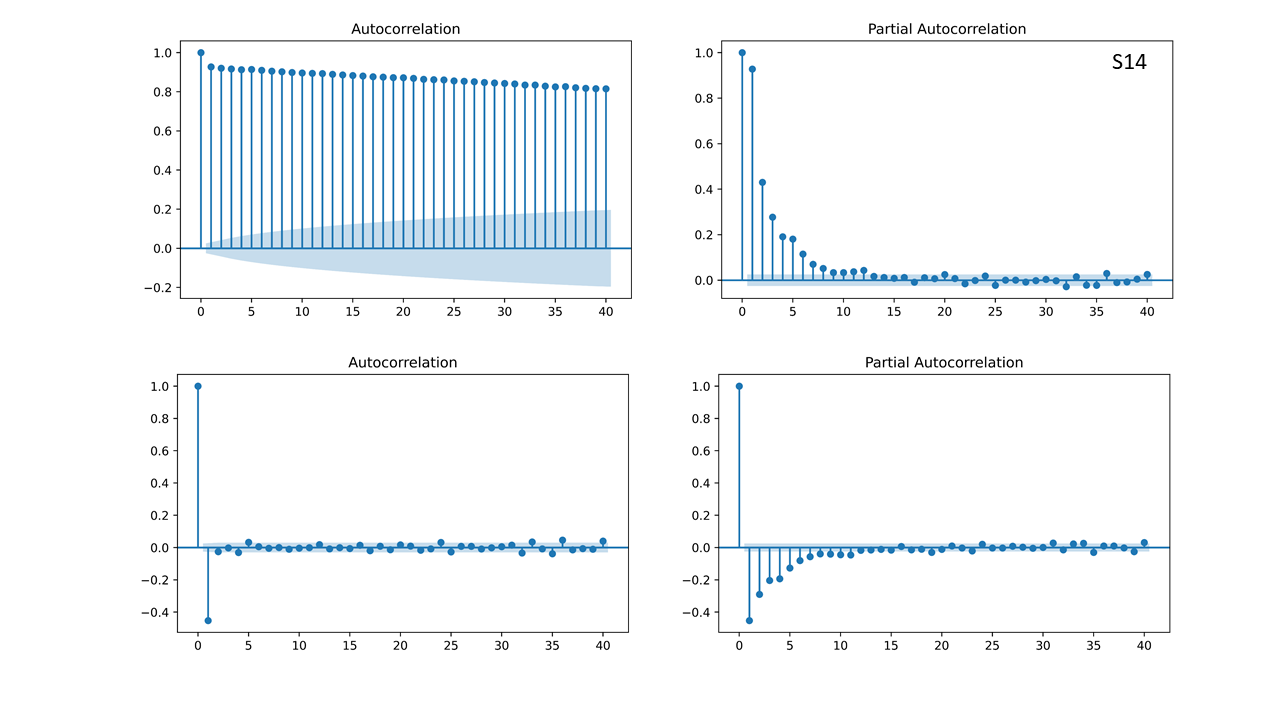}
\caption{ACF and PCF plots calculated form the residuals (top) and first-difference residuals plot (bottom) for S14 are shown.   }
\end{figure}

 \bibliographystyle{elsarticle-num} 
 \bibliography{cas-refs}





\end{frontmatter}